\documentclass[a4paper,11pt]{article}
\pdfoutput=1 
\usepackage{jheppub}
\usepackage{amsmath}
\usepackage{amsfonts}
\usepackage{amssymb}
\usepackage{graphicx}%
\usepackage[T1]{fontenc} 
\newcommand{\dhd}{{\textstyle d}
\lower.03ex\hbox{\kern-0.38em$^{\scriptstyle-}$}\kern-0.05em{}}
\newcommand{\dbar}{{\textstyle \delta}
\lower.03ex\hbox{\kern-0.38em$^{\scriptstyle-}$}\kern-0.05em{}}
\newcommand{\half}{{1\over 2}}
\newcommand{\bu}{{\bullet}}

\newcommand{\bsi}{{\bar \psi}}
\newcommand{\Bsi}{{\bar \Psi}}

\newcommand{\bxi}{{\bar \xi}}

\newcommand{\barA}{{\bar A}}
\newcommand{\barB}{{\bar B}}
\newcommand{\barC}{{\bar C}}
\newcommand{\barD}{{\bar D}}
\newcommand{\barG}{{\bar G}} 
\newcommand{\barF}{{\bar F}}   
 
\newcommand{\barP}{{\bar P}}

\newcommand{\Bigma}{{\bar \Sigma}}

\newcommand{\cala}{{\cal A}}

\newcommand{\cald}{{\cal D}}  
\newcommand{\calf}{{\cal F}}
 
\newcommand{\kal}{{\cal L}}

\newcommand{\calo}{{\cal O}}    
  
\newcommand{\calp}{{\cal P}}

\newcommand{\calv}{{\cal V}} 
\newcommand{\calu}{{\cal U}}

\newcommand{\hatV}{{\hat V}} 
\newcommand{\hatU}{{\hat U}}

\newcommand{\hatp}{{\hat p}}

\newcommand{\hatA}{{\hat A}}
\newcommand{\hatB}{{\hat B}}
\newcommand{\hatF}{{\hat F}}

\newcommand{\hsi}{{\hat \psi}}

\newcommand{\notA}{{\not\! A}}
\newcommand{\notp}{{\not\! p}}

\newcommand{\tilA}{{\tilde A}}
\newcommand{\tilB}{{\tilde B}}
\newcommand{\tilC}{{\tilde C}}

\newcommand{\tilF}{{\tilde F}}

\newcommand{\tiL}{{\tilde L}}

\newcommand{\tilS}{{\tilde S}}

\newcommand{\tilU}{{\tilde U}}
\newcommand{\tilV}{{\tilde V}}

\newcommand{\tikal}{\tilde{\cal L}}

\newcommand{\tsi}{\tilde {\psi}} 
\newcommand{\tipsi}{\tilde {\psi}}

\newcommand{\mato}{\mathbb{O}}

\pdfoutput=1
\abstract{We calculate power corrections to TMD factorization for particle production 
by gluon-gluon fusion in hadron-hadron collisions.}
\arxivnumber{}
\affiliation{$^a$ Physics Department, Old Dominion University, Norfolk, VA 23529, USA and Thomas Jefferson National Accelerator Facility, Newport News, VA 23606, USA}
\affiliation{$^b$ Physics Department, Brookhaven National Laboratory, Upton, NY 11973, USA}
\emailAdd{balitsky@jlab.org}
\emailAdd{atarasov@bnl.gov}
\begin{document}

\title{\boldmath Higher-twist corrections to gluon TMD factorization}
\author{I. Balitsky$^a$ and A. Tarasov$^b$}
\preprint{BNL-113982-2017-JA, JLAB-THY-17-2484}
\maketitle

\flushbottom

\section{Introduction\label{aba:sec1}}

\bigskip

Particle production  in hadron-hadron scattering with transverse momentum of produced particle  much smaller 
than its invariant mass is described in the framework of TMD factorization \cite{Collins:2011zzd, Collins:1981uw, Collins:1984kg, Ji:2004wu, GarciaEchevarria:2011rb}.
The typical example is the Higgs production at LHC through gluon-gluon fusion. Factorization formula for particle production in hadron-hadron scattering  looks like \cite{Collins:2011zzd, Collins:2014jpa}
\begin{eqnarray}
&&\hspace{-2mm}
{d\sigma\over  d\eta d^2q_\perp}~=~\sum_f\!\int\! d^2b_\perp e^{i(q,b)_\perp}
\cald_{f/A}(x_A,b_\perp,\eta)\cald_{f/B}(x_B,b_\perp,\eta)\sigma(ff\rightarrow H)
\nonumber\\
&&\hspace{-2mm}
+~{\rm power ~corrections}~+~{\rm Y-terms}
\label{TMDf}
\end{eqnarray}
where $\eta$ is the rapidity, $\cald_{f/A}(x,z_\perp,\eta)$ is the TMD density of  a parton $f$  in hadron $A$, and $\sigma(ff\rightarrow H)$ is the cross section of production of particle $H$ 
of invariant mass $m_H^2=Q^2$ in the scattering of two partons.
(For simplicity,  we consider the scattering of unpolarized hadrons.)

In this paper we calculate the first power corrections $\sim {q_\perp^2\over Q^2}$  in a sense that we represent them as a TMD-like matrix elements of higher-twist operators. It should be noted that our method works for arbitrary relation between $s$ and $Q^2$ 
 and between $q_\perp^2$ and hadron mass $m^2$ (provided that pQCD is applicable), but in this paper we only present the result for 
 the physically interesting region $s\gg Q^2\gg q_\perp^2\gg m^2$.

To obtain  formula (\ref{TMDf}) with first corrections we use factorization in rapidity \cite{Balitsky:1998ya}. We denote quarks and gluons with rapidity close to the rapidity of the projectile and target protons
as $A$-fields and $B$-fields, respectively. We call the remaining fields in the central region of rapidity by the name $C$-fields and integrate over them in the corresponding
functional integral. At this step, we get the effective action depending on $A$ and $B$ fields. 
The  subsequent integration over $A$ fields gives matrix elements
of some TMD-like operators switched between projectile proton states and  integration over $B$ fields will give matrix elements between target states. 
\footnote{It should be noted that due to the kinematics $Q^2\gg Q_\perp^2,m^2$ we will not need the explicit form of the high-energy effective action which is 
much sought after in the small-x physics but not known  up to now
except a couple of first perturbative terms  \cite{Balitsky:1998ya, Balitsky:2005we, Hatta:2005rn, Hatta:2005ia, Altinoluk:2009je}.}

The paper is organized as follows.  In Sect.  \ref{sec:funt} we derive the TMD factorization
from the double functional integral for the cross section of particle production. 
In Sect 2, which is central to our approach, we explain the method of
calculation of higher-twist power corrections based on a solution of classical Yang-Mills equations.
In Sect. \ref{sec:lhtc} we find the leading higher-twist correction to
particle production in the region $s\gg Q^2\gg q_\perp^2$. Finally, in Sect. \ref{sec:sx} we compare our calculations
in the small-$x$ limit to the classical field resulting from the scattering of two shock waves. The Appendices contain proofs
of some necessary technical statements.

\section{TMD factorization from functional integral \label{sec:funt}}

We consider production of an (imaginary) scalar particle $\Phi$ in proton-proton scattering. 
This particle is connected to gluons by the vertex
\begin{equation}
{\kal}_\Phi~=~g_\Phi\!\int\! d^4x~\Phi(x)g^2F^2(x),~~~~~F^2(x)~\equiv~F^a_{\mu\nu}(x)F^{a\mu\nu}(x)
\end{equation}
%
\begin{figure}[htb]
\begin{center}
\includegraphics[width=131mm]{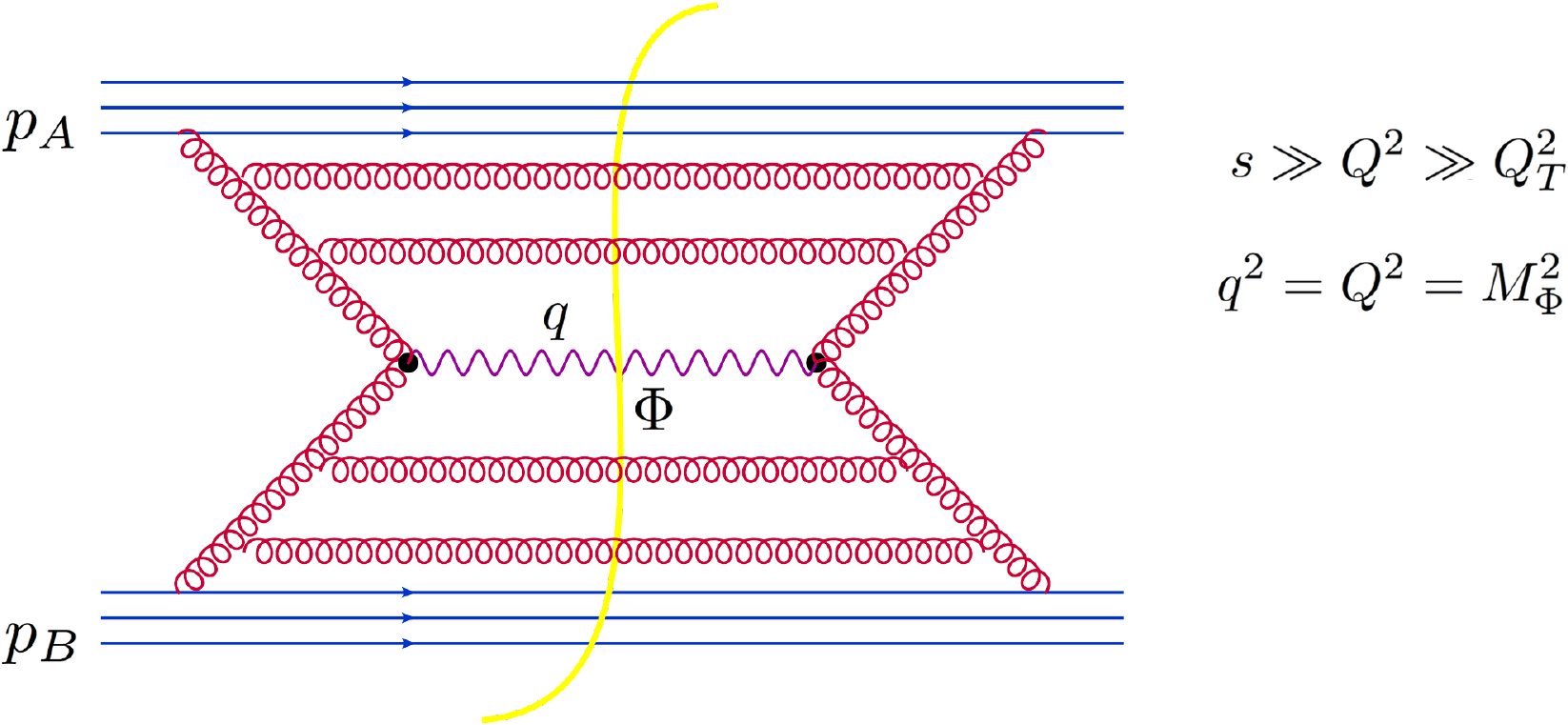}
\end{center}
\caption{Particle production by gluon-gluon fusion \label{fig:1}}
\end{figure}
This is a ${m_H\over m_t}\ll 1$ approximation \cite{Kniehl:1995tn, Shifman:1979eb}
 for Higgs production via gluon fusion at LHC
with 
$$
g_H~=~{1\over 48\pi^2 v}\big(1+{11\over 4\pi}\alpha_s+...\big)
$$
where $\alpha_s={g^2\over 4\pi}$ as usual. \footnote
{For finite $m_t$ the constant $g_H$ should be multiplied by ${3\tau \over 2}\big[1 + (1-\tau)\arcsin^2{1\over\sqrt\tau}\big]$ with $\tau={4m_t^2\over m_H^2}$ \cite{Gunion:1989we}.
} The differential cross section of $\Phi$ production has the form
\begin{eqnarray}
&&\hspace{-2mm}
d\sigma~=~{d^3q\over 2E_q(2\pi)^3}{g_\Phi^2\over 2s}W(p_A,p_B,q)
\end{eqnarray}
where we defined the ``hadronic tensor''  $W(p_A,p_B,q)$ as 
\begin{eqnarray}
\hspace{-1mm}
W(p_A,p_B,q)~&\stackrel{\rm def}{=}&~\sum_X\!\int\! d^4x~e^{-iqx}
\langle p_A,p_B|g^2F^2(x)|X\rangle\langle X|g^2F^2(0) |p_A,p_B\rangle
\nonumber\\
~&=&~\!\int\! d^4x~e^{-iqx}
\langle p_A,p_B|g^4F^2(x)F^2(0) |p_A,p_B\rangle  
\label{W}
\end{eqnarray}
As usual,  $\sum_X$ denotes the sum over full set of ``out''  states.  
It can be represented by double functional integral
\begin{eqnarray}
&&\hspace{-2mm}
W(p_A,p_B,q)~=~\sum_X\!\int\! d^4x~e^{-iqx}
\langle p_A,p_B|g^2F^2(x)|X\rangle\langle X|g^2F^2(0) |p_A,p_B\rangle
\label{dablfun}\\
&&\hspace{-2mm}
=\lim_{t_i\rightarrow -\infty}^{t_f\rightarrow\infty}g^4\!\!\int \! d^4x~ e^{-iqx}
\!\int^{\tilA(t_f)=A(t_f)}\!\!  D\tilA_\mu DA_\mu \!\int^{\tsi(t_f)=\psi(t_f)}\! D\tilde{\bar\psi}D\tilde{\psi} D\bsi D\psi 
~\Psi^\ast_{p_A}(\vec{\tilA}(t_i),\tipsi(t_i))
\nonumber\\
&&\hspace{-2mm}
\times~\Psi^\ast_{p_B}(\vec{\tilA}(t_i),\tipsi(t_i))e^{-iS_{\rm QCD}(\tilA,\tipsi)}e^{iS_{\rm QCD}(A,\psi)}
\tilF^2(x)F^2(0)\Psi_{p_A}(\vec{A}(t_i),\psi(t_i))\Psi_{p_B}(\vec{A}(t_i),\psi(t_i))
\nonumber
\end{eqnarray}
Here the fields $A,\psi$ correspond to the amplitude  $\langle X|F^2(0) |p_A,p_B\rangle$, fields $\tilA,\tipsi$ correspond to 
complex conjugate amplitude $\langle p_A,p_B|F^2(x)|X\rangle$
and  $\Psi_p(\vec{A}(t_i),\psi(t_i))$ denote the proton wave function at the initial time $t_i$. The boundary conditions
$\tilA(t_f)=A(t_f)$ and $\tsi(t_f)=\psi(t_f)$ reflect the sum over all states $X$, cf. Refs. \cite{Balitsky:1988fi}, \cite{Balitsky:1990ck}, \cite{Balitsky:1991yz}.

We use 
Sudakov variables $p=\alpha p_1+\beta p_2+p_\perp$ and the notations $x_\bu\equiv x_\mu p_1^\mu$ and $x_\ast\equiv x_\mu p_2^\mu$ 
for the dimensionless light-cone coordinates ($x_\ast=\sqrt{s\over 2}x_+$ and $x_\bu=\sqrt{s\over 2}x_-$). Our metric is $g^{\mu\nu}~=~(1,-1,-1,-1)$ so 
that $p\cdot q~=~(\alpha_p\beta_q+\alpha_q\beta_p){s\over 2}-(p,q)_\perp$ where $(p,q)_\perp\equiv -p_iq^i$. Throughout the paper, the sum over the Latin indices $i$, $j$... runs over the two transverse components while the sum over Greek indices runs over the four components as usual.

To derive the factorization formula, we separate the (quark and gluon) fields in the functional integral (\ref{dablfun}) into three sectors: ``projectile'' fields $A_\mu, \psi_a$ 
with $|\beta|<\sigma_a$, 
`` target'' fields with $|\alpha|<\sigma_b$ and ``central rapidity'' fields $C_\mu,\psi$ with $|\alpha|>\sigma_b$ and $|\beta|>\sigma_a$:
\footnote{The standard factorization scheme for particle production in hadron-hadron 
scattering is splitting  the diagrams in collinear to projectile part, collinear to 
target part, hard factor, and soft factor \cite{Collins:2011zzd}. Here we factorize only in rapidity. For
our purpose of calculation of power corrections in the tree approximation it is sufficient;
however, we hope to treat possible logs of transverse scales in loop corrections
in the same way as it was done in our rapidity evolution equations for gluon TMDs in Refs. \cite{Balitsky:2015qba, Balitsky:2016dgz}.}
\begin{figure}[htb]
\begin{center}
\includegraphics[width=151mm]{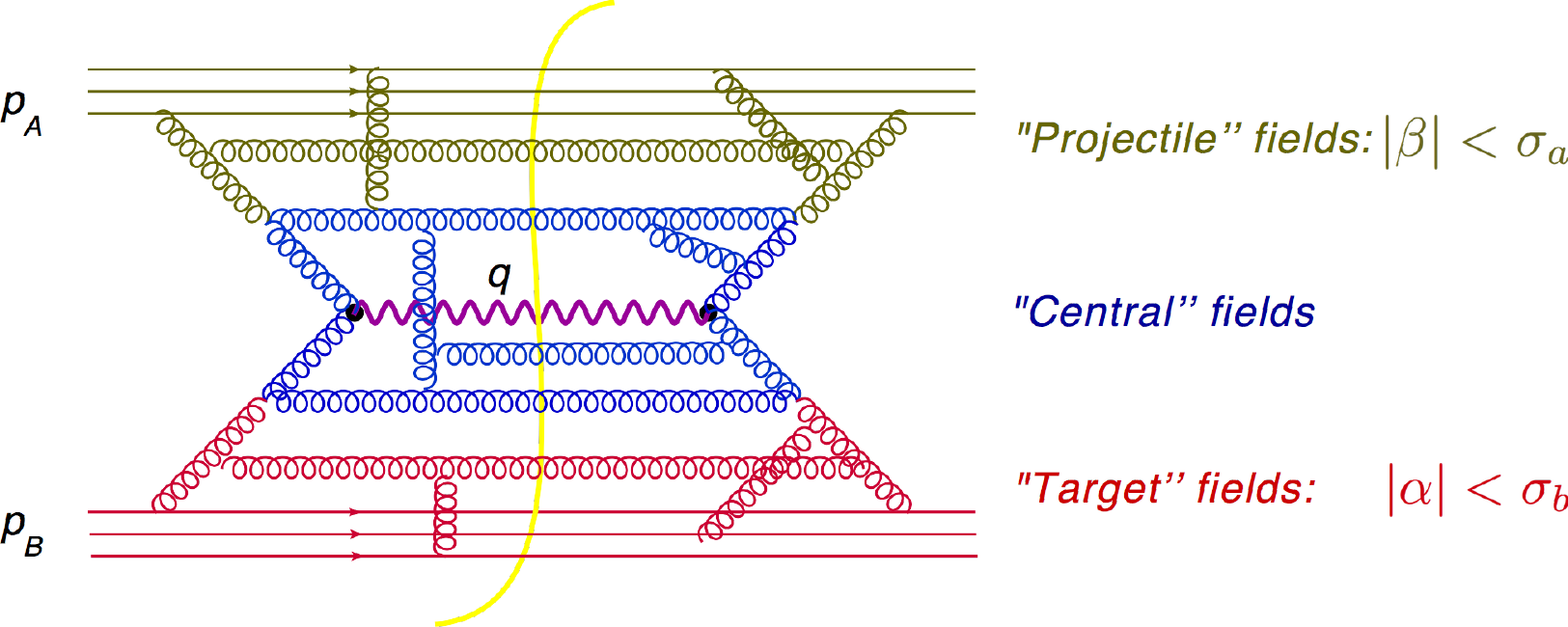}
\end{center}
\caption{Rapidity factorization for particle production \label{fig:2}}
\end{figure}
%
\begin{eqnarray}
&&\hspace{-1mm}
W(p_A,p_B,q)~=~g^4\!\int \! d^4x e^{-iqx}
\int^{\tilA(t_f)=A(t_f)}\! D\tilA_\mu DA_\mu
\int^{\tipsi_a(t_f)=\psi_a(t_f)}D\bsi_a D\psi_a D\tilde{\bar\psi}_aD\tilde{\psi}_a 
\nonumber\\
&&\hspace{-1mm}
\times 
~
e^{-iS_{\rm QCD}(\tilA,\tipsi_a)}e^{iS_{\rm QCD}(A,\psi_a)}\Psi^\ast_{p_A}(\vec{\tilA}(t_i),\tipsi_a(t_i))
\Psi_{p_A}(\vec{A}(t_i),\psi(t_i))
\nonumber\\
&&\hspace{-1mm}
\times
\int^{\tilB(t_f)=B(t_f)}\! D\tilB_\mu DB_\mu 
\int^{\tipsi_b(t_f)=\psi_b(t_f)}\!\!D\bsi_b D\psi_b D\tilde{\bar\psi}_bD\tilde{\psi}_b 
\nonumber\\
&&\hspace{-1mm}
\times~e^{-iS_{\rm QCD}(\tilB,\tipsi_b)}e^{iS_{\rm QCD}(B,\psi_b)}
\Psi^\ast_{p_B}(\vec{\tilB}(t_i),\tipsi_b(t_i))\Psi_{p_B}(\vec{B}(t_i),\psi_b(t_i))
\label{W2}\\
&&\hspace{-1mm}
\times~
\int\!DC_\mu\! \int^{\tilC(t_f)=C(t_f)} \! D\tilC_\mu \!\int\!D\bsi_C D\psi_C \int^{\tsi_c(t_f)=\psi_c(t_f)}\! D\tilde{\bar\psi}_C D\tsi_C~\tilF_C^2(x)F_C^2(0)~e^{-i\tilS_C+iS_C}
\nonumber
\end{eqnarray}
where $S_C=S_{\rm QCD}(A+B+C)-S_{\rm QCD}(A)-S_{\rm QCD}(B)$.

Our goal is to integrate over central fields and get
the amplitude in the factorized form, as a (sum of) products of functional integrals over $A$ fields representing projectile matrix elements (TMDs) 
and functional integrals over $B$ fields representing target matrix elements. In the spirit of background-field method, we ``freeze'' projectile and target fields (and denote them the $\barA$, $\bxi_a$, $\xi_a$ and $\barB$, $\bxi_b$, $\xi_b$ 
respectively)  and get a sum of diagrams in these external fields. 
Since  $|\beta|<\sigma_a$ in the projectile fields and $|\alpha|<\sigma_b$  in the target fields, at the  tree-level 
one can set with power accuracy $\beta=0$ for the  projectile fields and $\alpha=0$ for the target fields - the corrections will
be $O\big({m^2\over\sigma_a s}\big)$ and  $O\big({m^2\over\sigma_b s}\big)$.
Beyond the tree level, one should expect that the integration over $C$ fields will produce
the logarithms of the cutoffs $\sigma_a$ and $\sigma_b$ which will cancel with the corresponding
logs in gluon TMDs of the projectile and the target.

As usual, diagrams disconnected from the vertices $F^2(x)$ and $F^2(0)$ (``vacuum bubbles'' in external fields) exponentiate 
so the result has the schematic form
\begin{eqnarray}
&&\hspace{-1mm}
\int\!DC_\mu\! \int^{\tilC(t_f)=C(t_f)} D\tilC_\mu \!\int\!D\bsi_C D\psi_C \int^{\tsi_c(t_f)=\psi_c(t_f)}\! 
D\tilde{\bar\psi}_C D\tsi_C~g^4\tilF_C^2(x)F_C^2(0)~e^{-i\tilS_C+iS_C}
\nonumber\\
&&\hspace{-1mm}
=~e^{S_{\rm eff}(U,V,\tilU,\tilV)}
\calo(q,x;A,\tilA,\psi_a\tipsi_a;B,\tilB,\psi_b,\tipsi_b)
\label{intc}
\end{eqnarray}
where $\calo^{\mu\nu}(q,x;A, \psi_A;B, \psi_B)$ 
is a sum of diagrams connected to $\tilF^2(x)F^2(0)$. 
Since rapidities of central fields and $A$, $B$ fields are very different, one should expect the result of
integration over C-fields to be represented in terms of Wilson-line operators constructed form $A$ and $B$ fields. 

The effective action has the form
\begin{eqnarray}
&&\hspace{-1mm}
S_{\rm eff}(U,V,\tilU,\tilV)~=~2{\rm Tr}\int\! d^2x_\perp\big[-i\tilU_i\tilV^i+i U_iV^i
\label{effekshn}\\
&&\hspace{-1mm}
+~\big(\tikal_i(\tilU,\tilV)\tikal^i(\tilU,\tilV)
-2\tikal_i(\tilU,\tilV)\kal^i(U,V)+\kal^i(U,V)\kal^i(U,V)\big)
\ln\sigma_a\sigma_bs+O(\ln\sigma_a\sigma_bs)^2\big]
\nonumber
\end{eqnarray}
where Wilson lines $U$ are made from projectile fields
\begin{eqnarray}
&&\hspace{-1mm}
U(x_\perp)~=~[\infty p_2+x_\perp,-\infty p_2+x_\perp]^{A_\ast},~~~~~
U_i~=~U^\dagger i\partial_i U
\nonumber
\end{eqnarray}
and Wilson lines $V$ from target fields
\begin{eqnarray}
&&\hspace{-1mm}
V(x_\perp)~=~[\infty p_1+x_\perp,-\infty p_1+x_\perp]^{B_\bu},~~~~~
V_i~=~V^\dagger i\partial_i V
\nonumber
\end{eqnarray}
and similarly for $\tilU$ and $\tilV$ in the left sector. The explicit form of ``Lipatov vertices'' $L_i(U,V)$ is presented in 
\cite{Balitsky:2004rr}. Unfortunately, the effective action beyond the first two terms in (\ref{effekshn}) is unknown, but
we will demonstrate below that for our purposes we do not need the explicit form of the effective action.

After integration over $C$ fields the amplitude (\ref{dablfun}) can be rewritten as
\begin{eqnarray}
&&\hspace{-1mm}
W(p_A,p_B,q)~=~\!\int \! d^4x e^{-iqx}\!
\int^{\tilA(t_f)=A(t_f)}\! D\tilA_\mu DA_\mu
\int^{\tipsi_a(t_f)=\psi_a(t_f)}D\bsi_a D\psi_a D\tilde{\bar\psi}_aD\tilde{\psi}_a 
\nonumber\\
&&\hspace{-1mm}
\times 
~
e^{-iS_{\rm QCD}(\tilA,\tipsi_a)}e^{iS_{\rm QCD}(A,\psi_a)}\Psi^\ast_{p_A}(\vec{\tilA}(t_i),\tipsi_a(t_i))
\Psi_{p_A}(\vec{A}(t_i),\psi(t_i))
\nonumber\\
&&\hspace{-1mm}
\times
\int^{\tilB(t_f)=B(t_f)}\! D\tilB_\mu DB_\mu 
\int^{\tipsi_b(t_f)=\psi_b(t_f)}\!\!D\bsi_b D\psi_b D\tilde{\bar\psi}_bD\tilde{\psi}_b 
\nonumber\\
&&\hspace{-1mm}
\times~e^{-iS_{\rm QCD}(\tilB,\tipsi_b)}e^{iS_{\rm QCD}(B,\psi_b)}
\Psi^\ast_{p_B}(\vec{\tilB}(t_i),\tipsi_b(t_i))\Psi_{p_B}(\vec{B}(t_i),\psi_b(t_i))
\nonumber\\
&&\hspace{-1mm}
\times~
e^{S_{\rm eff}(U,V,\tilU,\tilV)}\calo(q,x;A,\psi_a,\tilA,\tipsi_a;B,\psi_b,\tilB,\tipsi_b)
\label{W3}
\end{eqnarray}
Note that due to boundary conditions at $t_f$ in the above  integral,   
the functional integral over $C$ fields in Eq. (\ref{intc}) should
be done in the background of the $A$ and $B$ fields satisfying 
\begin{equation}
\tilA(t_f)~=~A(t_f),~~~\tipsi_a(t_f)~=~\psi_a(t_f)~~{\rm and} ~~~\tilB(t_f)~=~B(t_f),~~~\tipsi_b(t_f)~=~\psi_b(t_f)
\label{baukon}
\end{equation}
Our approximation at the tree level is that $\beta=0$ for $A,\tilA$ fields and $\alpha=0$ for $B,\tilB$ fields 
which corresponds to  $A=A(x_\bu,x_\perp),~\tilA=\tilA(x_\bu,x_\perp)$ and $B=B(x_\ast,x_\perp),~\tilB=\tilB(x_\ast,x_\perp)$.

Now comes the important point: because of boundary conditions (\ref{baukon}),
for the purpose of calculating the integral (\ref{intc})  over central fields one can set 
\begin{eqnarray}
&&
A(x_\bu,x_\perp)=\tilA(x_\bu,x_\perp),~~~~\psi_a(x_\bu,x_\perp)=\tipsi_a(x_\bu,x_\perp)
\nonumber\\
&&
{\rm and}
\nonumber\\
&&
B(x_\ast,x_\perp)=\tilB(x_\ast,x_\perp),~~~~\psi_b(x_\ast,x_\perp)=\tipsi_b(x_\ast,x_\perp)
\label{baukond}
\end{eqnarray}
Indeed, because  $A,\psi$ and $\tilA,\tipsi$ do not depend on $x_\ast$, if they coincide at $x_\ast=\infty$ they should coincide everywhere.
Similarly, if $B,\psi_b$ and $\tilB,\tipsi_b$ do not depend on $x_\bu$, if they coincide at $x_\bu=\infty$ they should be equal.

It should be emphasized that the boundary conditions (\ref{baukon}) mean the summation over all intermediate states
in corresponding projectile and target matrix elements in the functional integrals over projectile and target fields. 
Without the sum over all intermediate states the conditions (\ref{baukond}) are no longer true.
For example, if we would like to measure another particle or jet in the fragmentation region of the projectile,  the second condition
in Eq. (\ref{baukond}) breaks down.
  
Next important observation is that due to Eqs. (\ref{baukond}) the effective action (\ref{effekshn}) vanishes 
for background fields satisfying conditions (\ref{baukon}).
For the first two terms displayed in (\ref{effekshn}) it is evident, but it is easy to see that the effective action in the 
background fields satisfying  (\ref{baukond}) should vanish due to unitarity. Indeed, let us consider the functional
integral (\ref{dablfun}) without sources $\tilde{F}^2(x)F^2(0)$. It describes the matrix element (\ref{crsc1}) without $\Phi$ production, that is 
\begin{eqnarray}
&&\hspace{-2mm}
\sum_X
\langle p_A,p_B|X\rangle
\langle X  |p_A,p_B\rangle  ~=~1
\label{crsc1}
\end{eqnarray}
(modulo appropriate normalization of $|p_A\rangle$ and $|p_B\rangle$ states). If we perform the same decomposition
into $A$, $B$, and $C$ fields as in Eq. (\ref{dablfun}) we will see integral (\ref{W3}) without $\calo^{\mu\nu}(q,x,y;A,\psi_a,\tilA,\tipsi_a;B,\psi_b,\tilB,\tipsi_b)$ 
which can be represented as
\begin{equation}
\langle p_A,p_B|e^{S_{\rm eff}(U,V,\tilU,\tilV)}|p_A,p_B\rangle~=~1
\end{equation}
which means that the effective action should vanish for the Wilson-line operators constructed from the  fields satisfying Eqs. (\ref{baukond}).
Summarizing, we see that at the tree level in our approximation
\begin{eqnarray}
&&\hspace{-1mm}
\int\!DC_\mu\! \int^{\tilC(t_f)=C(t_f)} D\tilC_\mu \!\int\!D\bsi_C D\psi_C \int^{\tsi_c(t_f)=\psi_c(t_f)}~D\tilde{\bar\psi}_C D\tsi_C~g^4\tilF_C^2(x)F_C^2(0)~e^{-i\tilS_C
+iS_C}
\nonumber\\
&&\hspace{-1mm}
=~
\calo(q,x;A,\psi_a;B,\psi_b)
\label{funtc}
\end{eqnarray}
where now $S_C~=~S_{\rm QCD}(C+A+B)-S_{\rm QCD}(A)-S_{\rm QCD}(B)$ and 
$\tilS_C~=~S_{\rm QCD}(\tilC+A+B)-S_{\rm QCD}(A)-S_{\rm QCD}(B)$. It is known that 
in the tree approximation the double functional integral (\ref{funtc}) is given by a set of 
retarded Green functions in the background fields \cite{Gelis:2003vh,Gelis:2006yv,Gelis:2007kn} (see also Appendix A for the proof).
Since the double functional integral (\ref{funtc}) is given by a set of retarded Green functions 
(in the background field $A+B$), the calculation of tree-level contributions to, say, $F^2(x)$  
in the r.h.s. of Eq. (\ref{funtc}) is equivalent to solving YM equation for 
$A_\mu(x)$ (and $\psi(x)$) with boundary conditions that the solution has the same asymptotics at $t\rightarrow -\infty$ 
as the superposition of incoming projectile and  target background fields. 

The hadronic tensor  (\ref{W3}) can now be represented as
\begin{eqnarray}
&&\hspace{-1mm}  
W(p_A,p_B,q)~=~\!\int \! d^4x e^{-iqx}
 \langle p_A|\langle p_B| \hat\calo(q,x;\hatA,\hsi_a;\hatB,\hsi_b)|p_A\rangle |p_B\rangle   
\label{W4}
\end{eqnarray}
where $\hat \calo(q,x;\hatA,\hsi_a;\hatB,\hsi_b)$ should be expanded in a series in $\hatA,\hsi_a;\hatB,\hsi_b$ operators
and evaluated between the corresponding (projectile or target) states: if
\begin{equation}
\hat\calo(q,x;\hatA,\hsi_a;\hatB,\hsi_b)
~=~\sum_{m,n}\! \int\! dz_mdz'_n c^{\mu\nu}_{m,n}(q,x)\hat\Phi_A(z_m)\hat\Phi_B(z'_n)
\end{equation}
(where $c^{\mu\nu}_{m,n}$ are coefficients and $\Phi$ can be any of $A_\mu$, $\psi$ or $\bsi$) then
\footnote{
Our logic here is the following: to get the expression
for $\hat\calo$ in Eq. (\ref{funtc}) we calculate $\calo$ in the background 
of two external fields $\Phi_A=(A_\mu,\psi_a)$ 
and $\Phi_B=(B_\mu,\psi_b)$ 
and then promote them to operators $\hat\Phi_A$ and $\hat\Phi_B$ in the obtained expressions for 
$\calo$. However, 
there is a subtle point in the promotion of background fields to operators. 
 When we are calculating 
$\calo$ as the r.h.s. of Eq. (\ref{funtc}) the fields $\Phi_A$ and $\Phi_B$ are c-numbers;
on the other hand, after functional integration in Eq. (\ref{dablfun}) they become operators which must be time-ordered
in the right sector and anti-time-ordered in the left sector. Fortunately, as we shall see below, all these operators are separated 
either by space-like distances or light-cone distances so all of them (anti) commute and thus can be treated as $c$-numbers.}
\begin{equation}
\hspace{-1mm}  
W~=~\!\int \! d^4x  e^{-iqx}
\sum_{m,n}\! \int\! dz_m c^{\mu\nu}_{m,n}(q,x)
\langle p_A|\hat\Phi_A(z_m)|p_A\rangle\!\int\! dz'_n\langle p_B| \hat\Phi_B(z'_n)|p_B\rangle   
\label{W5}
\end{equation}
As we will demonstrate below, the relevant operators are  quark and gluon fields with Wilson-line type gauge links 
collinear to either $p_2$ for $A$ fields or  $p_1$ for $B$ fields.

\section{Power corrections and solution of classical YM equations \label{sect:power}}
\subsection{Power counting for background fields}
As we discussed in previous Section, to get the hadronic tensor in the form  (\ref{W4}) we need to calculate 
the functional integral (\ref{funtc}) in the background of the fields (\ref{baukond}). 
 To understand the relative 
strength of Lorentz components of these fields, let us compare the typical term in the leading contribution to $W$ 
\begin{eqnarray}
&&\hspace{-1mm}
{64/s^2\over N_c^2-1}\!\int\! d^4x~e^{-iqx}
\langle p_A|\hatU_\ast^{mi}(x_\bu,x_\perp)\hatU_\ast^{mj}(0)|p_A\rangle
\langle p_B|\hatV_{\bu i}^n(x_\ast,x_\perp)\hatV_{\bu j}^n(0)|p_B\rangle
\end{eqnarray}
where 
\begin{equation}
\hatU^a_{\ast i}(z_\bu,z_\perp)~\equiv~[-\infty_\bu,z_\bu]_z^{ab}
g\hatF^b_{\ast i}(z_\bu,z_\perp),~~~~~
\hatV^a_{\bu i}(z_\ast,z_\perp)~\equiv~ [-\infty_\ast,z_\ast]_z^{ab}g\hatF^b_{\bu i}(z_\ast,z_\perp)
\end{equation}
and some typical higher-twist terms. 
As we mentioned, we consider $W(p_A,p_B,q)$ in the region where $s,Q^2\gg Q_\perp^2,m^2$ 
while the relation between $Q_\perp^2$ and $m^2$ and between $Q^2$ and $s$ may be arbitrary. 
So, for the purpose of counting of powers of $s$, we will not distinguish between $s$ and $Q^2$ 
(although at the final step we will be able to tell the difference since our final expressions for 
higher-twist corrections will have either $s$ or $Q^2$ in denominators). Similarly, for the 
purpose of power counting we will not distinguish between $m$ and $Q_\perp$ and will 
introduce $m_\perp$ which may be of order of $m$ or $Q_\perp$ depending on matrix element.

The estimate of the leading-twist matrix element between projectile states is
\begin{equation}
\hspace{-1mm}
\langle p_A|\hatU_{\ast i}^{a}(x_\bu,x_\perp)\hatU_{\ast j}^{a}(0)|p_A\rangle
~=~p_2^\mu p_2^\nu~\langle p_A|\hatU_{\mu i}^{a}(x_\bu,x_\perp)\hatU_{\nu j}^{a}(0)|p_A\rangle
\sim s^2\big(m_\perp^2g^\perp_{ij}+ m_\perp^4x^\perp_ix^\perp_j\big)
\label{3.3}
\end{equation}
(here we assume normalization $\langle p_A|p_A\rangle~=~1$ for simplicity).

The typical higher-twist correction is proportional to (see e.g. Eq. (\ref{Wmain}))
\begin{eqnarray}
&&\hspace{-1mm}
d^{abc}\langle p_A|\hatU_{\ast i}^{a}(x_\bu,x_\perp)\hatU_{\ast k}^{b}(x'_\bu,x_\perp)\hatU_{\ast j}^{c}(0)|p_A\rangle
\nonumber\\
&&\hspace{-1mm}
=~d^{abc}p_2^\mu p_2^\nu p_2^\lambda\langle p_A|\hatU_{\mu i}^{a}(x_\bu,x_\perp)
\hatU_{\nu k}^{b}(x'_\bu,x_\perp)\hatU_{\lambda j}^{c}(0)|p_A\rangle
\nonumber\\
&&\hspace{-1mm}
\sim~s^3m_\perp^4\big(g^\perp_{ij}x_k+
g^\perp_{ik}x_j
+g^\perp_{jk}x_i\big)~+~s^3m_\perp^6x_ix_jx_k
\end{eqnarray}

Since $x^\perp_i\sim {q^\perp_i\over q_\perp^2}\sim {1\over m_\perp}$ we see  that an extra 
$\hatF_{\mu i}$ in the matrix element between projectile states  brings 
$p_{1\mu}m_\perp$ 
which means that $\hatU_{\ast i}\sim sm_\perp$.

 Next,
some of the higher-twist matrix elements have an extra $U_{kl}$ like
\begin{equation}
\hspace{-0mm}
d^{abc}\langle p_A|\hatU_\ast^{ai}(x_\bu,x_\perp)\hatU^b_{kl}(x'_\bu,x_\perp)\hatU_\ast^{cj}(0)|p_A\rangle
\end{equation}
where 
\begin{equation}
\hatU_{kl}(x_\bu,x_\perp)~\equiv~
[-\infty_\bu,x_\bu]_xg\hatF_{kl}(x_\bu,x_\perp)[x_\bu,-\infty_\bu]_x
\label{ukl}
\end{equation}
Since we consider only unpolarized projectile and target hadrons
\begin{eqnarray}
&&\hspace{-1mm}
d^{abc}\langle p_A|\hatU_\ast^{ai}(x_\bu,x_\perp)\hatU_{kl}^{b}(x'_\bu,x_\perp)\hatU_\ast^{cj}(0)|p_A\rangle
\nonumber\\
&&\hspace{-1mm}
\sim~s^2\big(m_\perp^4 g^\perp_{ik}g^\perp_{jl}+m_\perp^6g^\perp_{ik} x_jx_l
+m_\perp^6 g^\perp_{jl}x_ix_k~-~k\leftrightarrow l\big)
\end{eqnarray}
and, comparing this to Eq. (\ref{3.3}), we see that  an extra $\hatF_{kl}$ can bring an extra  $m_\perp^2$. 
 Combining this with an estimate $U_{\ast i}\sim sm_\perp$ we see that the typical field 
 $\barA_\ast$ is of order $s$ while $\barA_i\sim m_\perp$. Similarly, for the target fields we get 
$\barB_\bu\sim s$, $\barB_i\sim m_\perp$. 

Some of the power corrections involve matrix elements like 
\begin{equation}
\hspace{-0mm}
d^{abc}\langle p_A|\hatU_\ast^{ai}(x_\bu,x_\perp)\hatU^b_{\ast\bu}(x'_\bu,x_\perp)\hatU_\ast^{cj}(0)|p_A\rangle
\end{equation}
where 
\begin{equation}
\hspace{-0mm}
\hatU_{\ast\bu}(x_\bu,x_\perp)~\equiv~
[-\infty_\bu,x_\bu]_xg\hatF_{\ast \bu}(x_\bu,x_\perp)[x_\bu,-\infty_\bu]_x
\label{ubu}
\end{equation}
An extra field strength operator $\hatF^{\mu\nu}$ between the projectile states can bring 
${p_A^\mu p_2^\nu\over p_A\cdot p_2}-\mu\leftrightarrow\nu$ so that $\hatF_{\ast\bu}\sim s m^2$
\footnote{The denominator $p_A\cdot p_2$ is due to the fact that $p_2$ enters only through the direction of Wilson line 
and therefore the matrix element should not change under rescaling $p_2\rightarrow\lambda p_2$}. 
Since $\barA_\ast\sim s$ we see that $\barA_\bu\sim m^2_\perp$. 
Similarly, for the target we get $\barB_\ast\sim m_\perp^2$.

Summarizing, the relative strength of the background gluon fields in projectile and target is
\begin{eqnarray}
&&\hspace{-1mm}
\barA_\ast(x_\bu,x_\perp)~\sim~s,~~~\barA_\bu(x_\bu,x_\perp)~\sim m_\perp^2,~~~~
\barA_i(x_\bu,x_\perp)~\sim~m_\perp
\nonumber\\
&&\hspace{-1mm}
\barB_\ast(x_\ast,x_\perp)~\sim~m_\perp^2,~~~\barB_\bu(x_\ast,x_\perp)~\sim s,~~~~
\barB_i(x_\ast,x_\perp)~\sim~m_\perp
\label{gluonfildz}
\end{eqnarray}

To finish power counting, we need also the relative strength of quark background fields $\psi_a$ and $\psi_b$. 
From classical equations for projectile and target
\begin{eqnarray}
&&\hspace{-1mm}
\barD^\mu\barA^a_{\mu\bu}~=~-g\bsi_a\gamma_\bu t^a\psi_a,~~~~\barD^\mu\barA^a_{\mu i}~=~-g\bsi_a\gamma_i t^a\psi_a,~~~~~~
\barD^\mu\barA^a_{\mu\ast}~=~-g\bsi_a\gamma_\ast t^a\psi_a
\nonumber\\
&&\hspace{-1mm}
\Big[{2\over s}(i\partial_\ast+g\barA_\ast)\hatp_1+{2g\over s}\barA_\bu\hatp_2
+(i\partial_i+g\barA_i)\gamma^i\Big]\psi_a~=~0
\nonumber\\
&&\hspace{-1mm}
\barD^\mu\barB^a_{\mu\bu}~=~-g\bsi_b\gamma_\bu t^a\psi_b,~~~~\barD^\mu\barB^a_{\mu i}~=~-g\bsi_b\gamma_i t^a\psi_b,~~~~~~
\barD^\mu\barB^a_{\mu\ast}~=~-g\bsi_b\gamma_\ast t^a\psi_b
\nonumber\\
&&\hspace{-1mm}
\Big[{2\over s}(i\partial_\bu+g\barB_\bu)\hatp_2+{2g\over s}\barB_\ast\hatp_1
+(i\partial_i+g\barB_i)\gamma^i\Big]\psi_b~=~0
\label{yds}
\end{eqnarray}
we get
\begin{eqnarray}
&&\hspace{-1mm}
\hatp_1\psi_a(x_\bu,x_\perp)~\sim~m_\perp^{5/2}, ~~~\gamma_i\psi_a(x_\bu,x_\perp)~\sim~m_\perp^{3/2}, ~~~~~
\hatp_2\psi_a(x_\bu,x_\perp)~\sim~s\sqrt{m_\perp}
\nonumber\\
&&\hspace{-1mm}
\hatp_1\psi_b(x_\ast,x_\perp)~\sim~s\sqrt{m_\perp}, ~~~\gamma_i\psi_b(x_\ast,x_\perp)~\sim~m_\perp^{3/2}, ~~~~~
\hatp_2\psi_b(x_\ast,x_\perp)~\sim~m_\perp^{5/2}
\label{kvarkfildz}
\end{eqnarray}
Thus, to find TMD factorization at the tree level (with higher-twist corrections) we need to calculate the functional integral (\ref{dablfun}) in the background fields of the strength given by Eqs. (\ref{gluonfildz}) and (\ref{kvarkfildz}).

\subsection{Approximate solution of classical equations}
\begin{figure}[htb]
\begin{center}
\includegraphics[width=131mm]{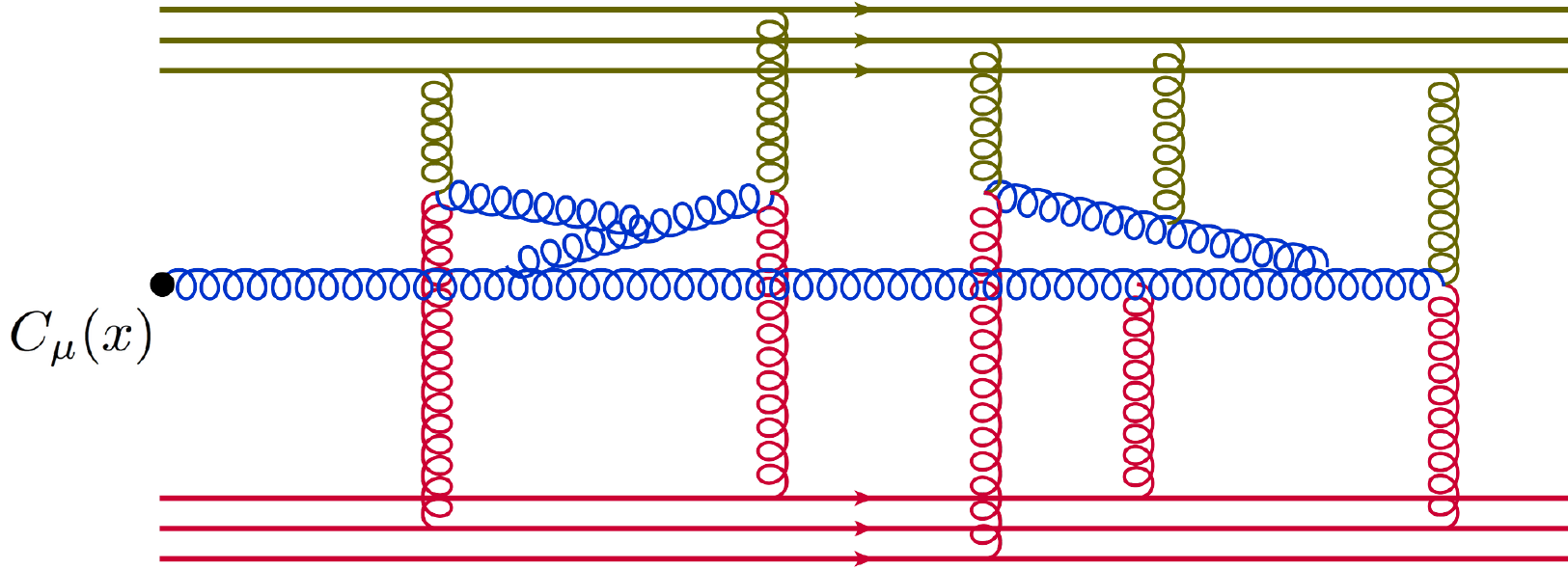}
\end{center}
\caption{Typical diagram for the classical field with projectile/target sources. The Green functions of the central fields are given by retarded propagators.   \label{fig:3}}
\end{figure}

As we discussed in Sect \ref{sec:funt}, the calculation of the functional integral (\ref{funtc}) over $C$-fields 
in the tree approximation reduces to finding fields $C_\mu$ and $\psi_c$ as solutions of Yang-Mills equations for the action
 $S_C~=~S_{\rm QCD}(C+A+B)-S_{\rm QCD}(A)-S_{\rm QCD}(B)$
\begin{eqnarray}
&&\hspace{-1mm}
D^\nu F_{\mu\nu}^a(\barA+\barB+C)~=~g\sum_f(\bsi^f_a+\bsi^f_b+\bsi^f_c)\gamma_\mu t^a(\psi^f_a+\psi^f_b+\psi^f_c)
\nonumber\\
&&\hspace{-1mm}
(i\not\!\partial+g\not\!\barA+g\not\!\barB+g\not\!C)(\psi^f_a+\psi^f_b+\psi^f_c)~=~m(\psi^f_a+\psi^f_b+\psi^f_c)
\label{yd}
\end{eqnarray}
As we discussed above (see also Appendix A) the solution of Eq. (\ref{yd}) which we need corresponds to the sum of set of diagrams
in background field $\barA+\barB$ with {\it retarded} Green functions (see Fig. \ref{fig:3}).
 The retarded Green
functions (in the background-Feynman gauge) are defined as 
\begin{eqnarray}
&&\hspace{-1mm}
(x|{1\over \barP^2g^{\mu\nu}+2ig\barF^{\mu\nu}+i\epsilon p_0}|y)
~\equiv~(x|{1\over p^2+i\epsilon p_0}|y)
-g(x|{1\over p^2+i\epsilon p_0}\calo_{\mu\nu}{1\over p^2+i\epsilon p_0}|y)
\nonumber\\
&&\hspace{-1mm}
+~g^2(x|{1\over p^2+i\epsilon p_0}\calo_{\mu\xi}
{1\over p^2+i\epsilon p_0}\calo^\xi_{~\nu}{1\over p^2+i\epsilon p_0}|y)
+...
\end{eqnarray}
where
\begin{eqnarray}
&&\hspace{-1mm}
\barP_\mu~\equiv~i\partial_\mu+g\barA_\mu+g\barB_\mu, ~~~~
\barF_{\mu\nu}~=~\partial_\mu(\barA+\barB)_\nu-\mu\leftrightarrow\nu-ig[\barA_\mu+\barB_\mu,\barA_\nu+\barB_\nu]
\nonumber\\
&&\hspace{-1mm}
\calo_{\mu\nu}~\equiv~\big(\{p^\xi,\barA_\xi+\barB_\xi\}+g(\barA+\barB)^2\big)g_{\mu\nu}+2i\barF_{\mu\nu}
\label{3.4}
\end{eqnarray}
and similarly for quarks.

The solutions of Eqs. (\ref{yd}) in terms of retarded Green functions give fields $C_\mu$ and $\psi_c$ that vanish at $t\rightarrow -\infty$. Thus, we are solving the usual classical YM equations
\begin{equation}
D^\nu F^a_{\mu\nu}~=~\sum_fg\bsi^f t^a\gamma_\mu\psi^f,~~~~(\not\!P-m_f)\psi^f~=~0
\label{kleqs}
\end{equation}
with boundary conditions
\begin{eqnarray}
&&\hspace{-11mm}
A_\mu(x)\stackrel{x_\ast\rightarrow -\infty}{=}\barA_\mu(x_\bu,x_\perp),~~~~
\psi(x)\stackrel{x_\ast\rightarrow -\infty}{=}\psi_a(x_\bu,x_\perp)
\nonumber\\
&&\hspace{-11mm}
A_\mu(x)\stackrel{x_\bu\rightarrow -\infty}{=}\barB_\mu(x_\ast,x_\perp),~~~~
\psi(x)\stackrel{x_\bu\rightarrow -\infty}{=}\psi_b(x_\ast,x_\perp)
\label{inicondi}
\end{eqnarray}
following from $C_\mu,\psi_c\stackrel{t\rightarrow -\infty}{\rightarrow} 0$.
These boundary conditions reflect the fact that at $t\rightarrow -\infty$ we have only incoming hadrons with ``A'' and ``B'' fields.

The solution of YM equations (\ref{kleqs}) in general case  is yet unsolved problem,
 especially important for scattering of two heavy nuclei
in semiclassical approximation. Fortunately, for our case of particle production with ${q_\perp\over Q}\ll 1$ we can construct the approximate solution of  (\ref{kleqs}) as a series in this small parameter. However, before doing this,
it is convenient to perform a gauge transformation so that the incoming projectile and target fields will no longer have large 
components $\sim s$ as $\barA_\ast$ and $\barB_\bu$  in Eq. (\ref{gluonfildz}). 
Let us perform the gauge transformation of Eq. (\ref{kleqs}) and initial conditions (\ref{inicondi}) with the gauge matrix $\Omega(x)$ such that
\begin{equation}
\hspace{-1mm}
\Omega(x_\ast,x_\bu,x_\perp)~\stackrel{x_\ast\rightarrow -\infty}{\rightarrow}~[x_\bu,-\infty_\bu]_x^{\barA_\ast},~~~~~
\Omega(x_\ast,x_\bu,x_\perp)~\stackrel{x_\bu\rightarrow -\infty}{\rightarrow}~[x_\ast,-\infty_\ast]_x^{\barB_\bu}
\label{omegasy}
\end{equation}
The existence of such matrix is proved in Appendix B by explicit construction. After such gauge transformation, 
the YM equation of course stays the same but the initial conditions (\ref{inicondi}) turn to
\begin{eqnarray}
&&\hspace{-11mm}
gA_\mu(x)\stackrel{x_\ast\rightarrow -\infty}{=}U_\mu(x_\bu,x_\perp),~~~~
\psi(x)\stackrel{x_\ast\rightarrow -\infty}{=}\Sigma_a(x_\bu,x_\perp)
\nonumber\\
&&\hspace{-1mm}
gA_\mu(x)\stackrel{x_\bu\rightarrow -\infty}{=}V_\mu(x_\ast,x_\perp),~~~~
\psi(x)\stackrel{x_\bu\rightarrow -\infty}{=}\Sigma_b(x_\ast,x_\perp)
\label{nuslo}
\end{eqnarray}
where 
\begin{eqnarray}
&&\hspace{-0mm}
U_\mu(x_\bu,x_\perp)~\equiv~{2\over s}p_{2\mu}U_\bu(x_\bu,x_\perp)+U_{\mu_\perp}(x_\bu,x_\perp)~~~~~
\label{uiv}\\
&&\hspace{-1mm}
V_\mu(x_\ast,x_\perp)~\equiv~{2\over s}p_{1\mu}V_\ast(x_\ast,x_\perp)+V_{\mu_\perp}(x_\ast,x_\perp)
\nonumber\\
&&\hspace{-1mm}
U_i(x_\bu,x_\perp)~\equiv~{2\over s}\!\int_{-\infty}^{x_\bu}\!dx'_\bu~U_{\ast i}(x'_\bu,x_\perp),~~~~~~
V_i(x_\ast,x_\perp)~\equiv~{2\over s}\!\int_{-\infty}^{x_\ast}\!dx'_\ast~V_{\bu i}(x'_\ast,x_\perp)
\nonumber\\
&&\hspace{-1mm}
U_\bu(x_\bu,x_\perp)~\equiv~{2\over s}\!\int_{-\infty}^{x_\bu}\!dx'_\bu~U_{\ast \bu}(x'_\bu,x_\perp),~~~~~~
V_\ast(x_\ast,x_\perp)~\equiv~-{2\over s}\!\int_{-\infty}^{x_\ast}\!dx'_\ast~V_{\ast\bu }(x'_\ast,x_\perp)
\nonumber
\end{eqnarray}

and $\Sigma_a, \Sigma_b$ are defined as
\begin{eqnarray}
&&\hspace{-0mm}
\Sigma_a(z_\bu, z_\perp)~\equiv~ [-\infty_\bu,z_\bu]_z\psi_a(z_\bu,z_\perp),~~~~
\Sigma_b(z_\ast, z_\perp)~\equiv~ [-\infty_\ast,z_\ast]_z\psi_b(z_\ast,z_\perp)
\label{sigmi}
\end{eqnarray}

The initial conditions (\ref{nuslo}) look like the projectile fields in the light-like gauge $p_2^\mu A_\mu=0$ and target
fields in the light-like gauge $p_1^\mu A_\mu=0$ so our construction of matrix $\Omega$ in a way proves
that we can take the sum of projectile fields in one gauge and target fields in another gauge as a zero-order 
approximation for iterative solution of the YM equations.  Note also that our power counting discussed in previous Section
means that 
\begin{equation}
\hspace{-1mm}
U_\bu~\sim~V_\ast~\sim~m_\perp^2,~~~~~~U_i~\sim~V_i~\sim m_\perp
\label{uvs}
\end{equation}
so we do not have large background fields $\sim s$ after this gauge transformation. Finally, the classical equations 
for projectile and target fields in this gauge read
\footnote{Here we consider only $u$, $ d$, and $s$ quarks which can be regarded as massless.}:
\begin{eqnarray}
&&\hspace{-1mm}
D_U^\nu U_{\mu\nu}^a~=~g^2\sum_f\Bigma^f_a\gamma_\mu t^a\Sigma^f_a,~~~i\!\not\! D_U\Sigma_a~=~0
\nonumber\\
&&\hspace{-1mm}
D_V^\nu V_{\mu\nu}^a~=~g^2\sum_f\Bigma^f_b\gamma_\mu t^a\Sigma^f_b,~~~i\!\not\! D_V\Sigma_b~=~0
\label{YMs}
\end{eqnarray}
where $U_{\mu\nu}\equiv\partial_\mu U_\nu-\partial_\nu U_\mu-i[U_\mu,U_\nu]$, 
$D_U^\mu\equiv (\partial^\mu-i[U^\mu,)$ and similarly for $V$ fields. 

We will solve Eqs. (\ref{kleqs}) iteratively,  order by order in perturbation theory,  starting from the 
zero-order approximation in the form of the sum of projectile and target fields
\begin{eqnarray}
&&\hspace{-1mm}
g\cala_\mu^{[0]}(x)~=~U_\mu(x_\bu,x_\perp)+V_\mu(x_\ast,x_\perp)
\nonumber\\
&&\hspace{-1mm}
\Psi^{[0]}(x)~=~\Sigma_a(x_\bu,x_\perp)+\Sigma_b(x_\ast,x_\perp)
\label{trials}
\end{eqnarray}
and improving it by calculation of Feynman diagrams with retarded propagators in the background fields (\ref{trials}).

The first step is the calculation of the linear term for the trial configuration (\ref{trials}). We rewrite field strength components as
\begin{eqnarray}
\hspace{-1mm}
g\calf^{[0]}_{\bu i}~=~U_{\bu i}+V_{\bu i}-i[U_\bu,V_i],&~~~~& g\calf^{[0]}_{\ast i}~=~U_{\ast i}+V_{\ast i}-i[V_\ast,U_i]~~
\\
\hspace{-1mm} 
g\calf^{[0]}_{\ast \bu}~=~U_{\ast \bu}+V_{\ast \bu}+i[U_\bu,V_\ast],&~~~&g\calf^{[0]}_{ij}~=~U_{ij}+V_{ij}-i[U_i,V_j]+i[U_j,V_i]
\nonumber
\end{eqnarray}
Note that $U_{\ast i}\sim V_{\bu i}\sim sm_\perp$, $U_{\ast \bu}\sim V_{\ast\bu}\sim sm_\perp^2$ while all other components are not large.

The linear term has the form
\begin{eqnarray}
&&\hspace{-1mm}
L^a_i~\equiv~\cald^\mu\calf_{\mu i}^{[0]a}+g\Bsi^{[0]} \gamma_it^a\Psi^{[0]}~=~L^{(0)a}_i+L^{(1)a}_i
\nonumber\\
&&\hspace{-1mm}
L^{(0)a}_i~=~
-{i\over g}\big[U^{jab}V^b_{ji}+V^{jab}U^b_{ji}+\cald_j^{ab}(U^{jbc}V_i^c+V^{jbc}U_i^c)\big]
\nonumber\\
&&\hspace{33mm}
-~{2i\over gs}\big(U_{\ast \bu}^{ab}V^b_i-V_{\ast\bu}^{ab}U^b_i\big)
+g\Bigma_at^a\gamma_i\Sigma_b+g\Bigma_bt^a\gamma_i\Sigma_a
\nonumber\\
&&\hspace{-1mm}
L^{(1)a}_i~=~-{2i\over gs}\big[U_\bu^{ab}V_{\ast i}^b+V_\ast^{ab}U_{\bu i}^b
-i\{U_\bu, V_\ast\}^{ab}U_i^b-i\{V_\ast, U_\bu\}^{ab}V_i^b\big]
\nonumber\\
&&\hspace{-1mm}
L_\bu^a~\equiv~~\cald^\mu\calf_{\mu \bu}^{[0]a}+g\Bsi^{[0]} \gamma_\bu t^a\Psi^{[0]}
~=~L^{(-1)a}_\bu+L^{(0)a}_\bu+L^{(1)a}_\bu
,~~~~~
L^{(-1)a}_\bu~=~{i\over g}U^{jab}V_{\bu j}^b
\nonumber\\
&&\hspace{-1mm}
L^{(0)a}_\bu~=~{i\over g}V^{jab}U_{\bu j}^b
+{i\over g}\cald^{jab}U_\bu^{bc}V_j^c+g\Bigma_at^a\gamma_\bu\Sigma_b+g\Bigma_bt^a\gamma_\bu\Sigma_a
-{4i\over gs}U_\bu^{ab}V_{\ast\bu}^b
\nonumber\\
&&\hspace{-1mm}
L^{(1)a}_\bu~=~{2\over gs}(U_\bu U_\bu)^{ab}V_{\ast}^b
\nonumber\\
&&\hspace{-1mm}
L_\ast^a~\equiv~~\cald^\mu\calf_{\mu \ast}^{[0]a}+g\Bsi^{[0]} \gamma_\ast t^a\Psi^{[0]}
~=~L^{(-1)a}_\ast+L^{(0)a}_\ast+L^{(1)a}_\ast
,~~~~~
L^{(-1)a}_\ast~=~{i\over g}V^{jab}U_{\ast j}^b
\nonumber\\
&&\hspace{-1mm}
L^{(0)a}_\ast~=~{i\over g}U^{jab}V_{\ast j}^b
+{i\over g}\cald^{jab}V_\ast^{bc}U_j^c+g\Bigma_at^a\gamma_\ast\Sigma_b+g\Bigma_bt^a\gamma_\ast\Sigma_a
+{4i\over gs}V_\ast^{ab}U_{\ast\bu}^b
\nonumber\\
&&\hspace{-1mm}
L^{(1)a}_\ast~=~{2\over gs}(V_\ast V_\ast)^{ab}U_{\bu}^b
\nonumber\\
&&\hspace{-1mm}
L_\psi~\equiv~\not\!P \Psi^{[0]}~=~L_\psi^{(0)}+L_\psi^{(1)}~~~~
\nonumber\\
&&\hspace{-1mm}
L_\psi^{(0)}~=~\gamma^iU_i\Sigma_b+\gamma^iV_i\Sigma_a
,~~~~
L_\psi^{(1)}~=~{2\over s}\hatp_2 U_\bu\Sigma_b+{2\over s}\hatp_1 V_\ast\Sigma_a
\label{linterm}
\end{eqnarray}
where $\cald^j~\equiv~\partial^j-iU^j-iV^j$, $\cald_\bu=\partial_\bu-iU_\bu$, and $\cald_\ast=\partial_\ast-iV_\ast$.
The power-counting estimates for linear terms in Eq. (\ref{linterm}) are
\begin{equation}
\begin{array}{lll}
\hspace{-1mm}
L^{(0)}_i~\sim~m_\perp^3,~~~~&~L^{(1)}_i~\sim~{m_\perp^5\over s}~~~&~
\\
\hspace{-1mm}
L^{(-1)}_\bu\sim L^{(-1)}_\ast~\sim~sm_\perp^2,~~~~~~~&L^{(0)}_\bu\sim L^{(0)}_\ast~\sim~m_\perp^4,~~~~&~~~~
L^{(1)}_\bu\sim L^{(1)}_\ast~\sim~{m_\perp^6\over s}
\\
\hspace{-1mm}
L_\psi^{(0)}~\sim~m_\perp^{5/2},~~~~&L_\psi^{(1)}~\sim~{m^{9/2}\over s}&
\end{array}
\label{lintermvalues}
\end{equation}
Note that the order of perturbation theory is labeled by $(...)^{[n]}$ and the order of expansion in the parameter
${m_\perp^2\over s}$ by $(...)^{(n)}$.

With the linear term (\ref{linterm}), a couple of  first terms in perturbative series are
\begin{eqnarray}
&&\hspace{-1mm}
A_\mu^{[1]a}(x)~=~\int\!d^4z~(x|{1\over \calp^2g^{\mu\nu}+2ig\calf^{[0]\mu\nu}}|z)^{ab}L^{b\nu}(z)
\label{fields}\\
&&\hspace{-1mm}
A_\mu^{[2]a}(x)~=~g\int\!d^4z~\Big[
-~i(x|{1\over \calp^2g^{\mu\eta}+2ig\calf^{[0]\mu\eta}}\calp^\xi|z)^{aa'}f^{a'bc}A^{[1]b}_\xi A^{[1]c\eta}
\nonumber\\
&&\hspace{15mm}
+~(x|{1\over \calp^2g^{\mu\eta}+2ig\calf^{[0]\mu\eta}}|z)^{aa'}f^{a'bc}A^{[1]b\xi} 
(\cald_\xi A^{[1]c\eta}-\cald^\eta A^{[1]c}_\xi)\Big]
\nonumber
\end{eqnarray}
for gluon fields (in the background-Feynman gauge) and 
\begin{eqnarray}
&&\hspace{-1mm}
\Psi_f^{[1]}(x)~=~-\!\int\! d^4z~(x|{1\over \not\! \calp}|z)L_\psi(z),~~~~
\Psi_f^{[2]}(x)~=~-g\!\int\! d^4z~(x|{1\over \not\! \calp}|z)\notA^{[1]}(z)\Psi_f^{[0]}(z)
\nonumber\\
\label{quarkfildz}
\end{eqnarray}
for quarks where 
\begin{equation}
\calp_\bu~=~i\partial_\bu+U_\bu,~~~~\calp_\ast~=~i\partial_\ast+V_\ast,~~~
\calp_i~=~i\partial_i+U_i+V_i
\end{equation}
are operators in external zero-order fields (\ref{trials}).
Hereafter we use Schwinger's notations for propagators in external fields normalized according to
 $(x|F(p)|y)\equiv \int\!\dhd^4 p~e^{-ip(x-y)}F(p)$ . Moreover, when it will not lead to a confusion,
 we will use short-hand notation\\ ${1\over \calo}\calo'(x)~\equiv~\int\! d^4z~(x|{1\over\calo}|z)\calo'(z)$.
 Next iterations will give us a set of tree-level Feynman diagrams 
in the background field $U_\mu+V_\mu$ and $\Sigma_a+\Sigma_b$.

Let us consider the fields in the first order in perturbation theory:
\begin{eqnarray}
&&\hspace{-1mm}
A_\mu^{[1]}~=~{1\over \calp^2g^{\mu\nu}+2ig\calf^{[0]\mu\nu}}L^\nu
\label{fields1}\\
&&\hspace{13mm}
=~
{1\over [\{\alpha+{2\over s}V_\ast,\beta+{2\over s}U_\bu\}{s\over 2}
-(p+U+V)_\perp^2]g^{\mu\nu}+2ig\calf^{[0]\mu\nu}+i\epsilon p_0}L^\nu
\nonumber\\
&&\hspace{-1mm}
\Psi_f^{[1]}(x)~=~-{1\over \not\! \calp}L_\psi~=~-{(\alpha+{2\over s}V_\ast)\notp_1+(\beta+{2\over s}U_\bu)\notp_2+\not\!\calp_\perp
\over  \{\alpha+{2\over s}V_\ast,\beta+{2\over s}U_\bu\}{s\over 2}-(p+U+V)_\perp^2+i\epsilon p_0}L_\psi
\nonumber
\end{eqnarray}
Here $\alpha$, $\beta$, and $p_\perp$ are understood as differential operators
$\alpha=i{\partial\over\partial x_\bu}$, $\beta=i{\partial\over\partial x_\ast}$  and $p_i=i{\partial\over\partial x^i}$.

Now comes the central point of our approach. Let us expand quark and gluon propagators in powers of background fields, then 
we get a set of diagrams shown in Fig. \ref{fig:3}.
 The typical bare gluon propagator in Fig. \ref{fig:3} is
\begin{equation}
{1\over p^2+i\epsilon p_0}~=~{1\over\alpha\beta s-p_\perp^2+i\epsilon(\alpha+\beta)}
\label{gluonpropagator}
\end{equation}
 Since we do not consider loops of $C$-fields in this paper, the  transverse momenta 
in tree diagrams  are determined by further integration over projectile (``A'') and target (``B'') fields 
in Eq. (\ref{W3}) which converge on  either $q_\perp$ or $m$. On the other hand, the integrals over 
 $\alpha$ converge on either $\alpha_q$ or $\alpha\sim 1$ and similarly the characteristic $\beta$'s
 are either $\beta_q$ or $\sim 1$.
Since $\alpha_q\beta_qs=Q_\parallel^2\gg Q_\perp^2$, one can expand gluon and quark propagators 
in powers of ${p_\perp^2\over \alpha\beta s}$ 
\begin{eqnarray}
&&
{1\over p^2+i\epsilon p_0}~=~{1\over s(\alpha+i\epsilon)(\beta+i\epsilon)}
\Big(1
+{p_\perp^2/s\over(\alpha+i\epsilon)(\beta+i\epsilon)}+...\Big)
\label{propexpan}\\
&&
{\not\! p\over p^2+i\epsilon p_0}~=~{1\over s}\Big({\notp_1\over \beta+i\epsilon}
+{\notp_2\over \alpha+i\epsilon}+{\notp_\perp\over (\alpha+i\epsilon)(\beta+i\epsilon)}\Big)
\Big(1
+{p_\perp^2/s\over(\alpha+i\epsilon)(\beta+i\epsilon)}+...\Big)
\nonumber
\end{eqnarray}
The explicit form of operators ${1\over \alpha+i\epsilon}$, ${1\over \beta+i\epsilon}$, and 
${1\over (\alpha+i\epsilon)(\beta+i\epsilon)}$ is
\begin{eqnarray}
(x|{1\over \alpha+i\epsilon}|y)
&=&~{s\over 2}\!\int\! \dhd^2p_\perp\!\int\! {\dhd\alpha\over\alpha+i\epsilon}\dhd\beta
~e^{-i\alpha(x-y)_\bu-i\beta(x-y)_\ast+i(p,x-y)_\perp}
\nonumber\\
&&=~-i{s\over 2}(2\pi)^2\delta^{(2)}(x_\perp-y_\perp)
\theta(x_\bu-y_\bu)\delta(x_\ast-y_\ast)
\nonumber\\
(x|{1\over \beta+i\epsilon}|y)
&=&~{s\over 2}\!\int\! \dhd^2p_\perp\!\int\! \dhd\alpha{\dhd\beta\over\beta+i\epsilon}
~e^{-i\alpha(x-y)_\bu-i\beta(x-y)_\ast+i(p,x-y)_\perp}
\nonumber\\
&&=~-i{s\over 2}(2\pi)^2\delta^{(2)}(x_\perp-y_\perp)
\theta(x_\ast-y_\ast)\delta(x_\bu-y_\bu)
\nonumber\\
(x|{1\over (\alpha+i\epsilon)(\beta+i\epsilon)}|y)
&=&~{s\over 2}\!\int\! \dhd^2p_\perp\!\int\! {\dhd\alpha\over\alpha+i\epsilon}{\dhd\beta\over\beta+i\epsilon} 
~e^{-i\alpha(x-y)_\bu-i\beta(x-y)_\ast+i(p,x-y)_\perp}
\nonumber\\
&&=~-{s\over 2}(2\pi)^2\delta^{(2)}(x_\perp-y_\perp)
\theta(x_\ast-y_\ast)\theta(x_\bu-y_\bu)
\label{propexplicit}
\end{eqnarray}
After the expansion (\ref{propexpan}), the dynamics in the transverse space effectively becomes trivial: 
all background fields stand either at $x$ or at $0$. (This validates the reasoning in the footnote on page 3).

One may wonder why we do not cut the integrals in Eq. (\ref{propexplicit}) to $|\alpha|>\sigma_b$ and $|\beta|>\sigma_a$
according to the definition of $C$ fields in Sect. \ref{sec:funt}.
\footnote{
Such cutoffs for integrals over $C$ fields are introduced explicitly 
in the framework of soft-collinear effecive theory, see the review \cite{Rothstein:2016bsq}.} 
The reason is that in the diagrams like Fig. \ref{fig:3}
with retarded propagators (\ref{propexplicit}) one can 
shift the contour of integration over $\alpha$ and/or $\beta$  to the complex plane away to avoid the region of 
small $\alpha$ or $\beta$.
\footnote{This may be wrong if there is pinching of poles in the integrals over $\alpha$ or $\beta$ 
but we will see that in our integrals for the tree-level power corrections the pinching of poles never occurs. In the higher orders in perturbation theory the pinching does occur so one needs to formulate a subtraction program to avoid double counting.}

Note that the background fields are also smaller than typical $p_\parallel^2\sim s$. Indeed, from Eq. (\ref{uvs}) 
we see that $p_\bu={s\over 2}\beta\gg U_\bu\sim m^2$ ( because $\alpha\geq\alpha_q\gg {m^2\over s}$) and similarly
$p_\ast\gg V_\ast$. Also $(p_i+U_i+V_i)^2~\sim~q_\perp^2\ll p_\parallel^2$.  The only exception is the fields
$V_{\bu i}$ or $U_{\ast i}$ which are of order of $sm_\perp$ but we will see that effectively the expansion in powers of
these fields is cut at the second term with our accuracy.

\subsection{Twist expansion of classical gluon fields}

Now we expand the classical gluon fields  in powers of  ${p_\perp^2\over p_\parallel^2}\sim{m_\perp^2\over s}$. 
It is clear that for the leading higher-twist correction we need to take into account only the first two terms (\ref{fields}) of 
the perturbative expansion of classical field. The expansion (\ref{fields})  of gluon field $A_\bu$  takes the form 
\begin{eqnarray}
&&\hspace{-1mm}
A^{[0]}_\bu+A^{[1]}_\bu~=~A^{(0)}_\bu+A^{(1)}_\bu+O\big({m_\perp^6\over s^2})~\hspace{2mm}
\nonumber\\
&&\hspace{-1mm}
A^{(0)a}_\bu~=~A^{([1]0)a}_\bu+{1\over g}U^a_\bu~=~{1\over p_\parallel^2}L^{(-1)a}_\bu+{1\over g}U^a_\bu~=~{1\over g}U^a_\bu+{1\over 2g\alpha}U_j^{ab}V^{jb}
\nonumber\\
&&\hspace{-1mm}
A^{(1)a}_\bu~=~{1\over p_\parallel^2}L^{(0)a}_\bu
~+~{1\over 2gp_\parallel^2}\big((\{\alpha,U_\bu\}+\{\beta,V_\ast\}-\calp_\perp^2)V^j\big)^{ab}{1\over\alpha}U_j^b
-~ 2i{1\over p_\parallel^2}(V_\bu^{~i})^{ab} A^{(1)b}_i 
\nonumber\\
&&\hspace{-1mm}
+~{4i\over s}{1\over p_\parallel^2}(U_{\ast\bu}+V_{\ast\bu})^{ab}{1\over p_\parallel^2}L^{(-1)b}_\bu
-~{igf^{abc}\over\alpha s}A^{([1]0)b}_\ast A^{([1]0)c}_\bu
-~{1\over p_\parallel^2}A^{([1]0)ab}_\bu U_{j}^{bc}V^{c j}
\label{expanbu}
\end{eqnarray}
where 
\begin{eqnarray}
&&\hspace{-1mm}
A^{([1]0)a}_\bu \equiv~{1\over p_\parallel^2}L^{(-1)a}_\bu~=~{i\over 2\alpha g}f^{abc}U_j^{b}V^{cj},~~~~~~~
A^{([1]0)a}_\ast \equiv~{1\over p_\parallel^2}L^{(-1)a}_\ast~=~-{i\over 2\beta g}f^{abc}U_j^{b}V^{cj}
\nonumber\\
&&\hspace{-1mm}
\Rightarrow~\cald_\ast A^{([1]0)a}_\bu-\cald_\bu A^{([1]0)a}_\ast
~=~{s\over 2g}f^{abc}U_j^b V^{cj}~+~O(m_\perp^2)
\label{4.21}
\end{eqnarray}
Similarly, from Eq. (\ref{fields}) one obtains 
\begin{eqnarray}
&&\hspace{-1mm}
A^{[0]}_\ast+A^{[1]}_\ast~=~A^{(0)}_\ast+A^{(1)}_\ast+O\big({m_\perp^6\over s^2})~\hspace{2mm}
\nonumber\\
&&\hspace{-1mm}
A^{(0)a}_\ast~=~A^{([1]0)a}_\ast+{1\over g}V^a_\ast
~=~{1\over p_\parallel^2}L^{(-1)a}_\ast+{1\over g}V^a_\ast~=~{1\over g}V^a_\ast-{1\over 2g\beta}U_j^{ab}V^{jb}
\nonumber\\
&&\hspace{-1mm}
A^{(1)a}_\ast~=~{1\over p_\parallel^2}L^{(0)a}_\ast
~+~{1\over 2gp_\parallel^2}\big((\{\alpha,U_\bu\}+\{\beta,V_\ast\}-\calp_\perp^2)U^j\big)^{ab}{1\over\beta}V_j^b
-~ 2i{1\over p_\parallel^2}(U_\ast^{~i})^{ab} A^{(1)b}_i 
\nonumber\\
&&\hspace{-1mm}
-~{4i\over s}{1\over p_\parallel^2}(U_{\ast\bu}+V_{\ast\bu})^{ab}A^{([1]0)b}_\ast
+~{igf^{abc}\over \beta s}A^{([1]0)b}_\ast A^{([1]0)c}_\bu
-~{1\over p_\parallel^2}A^{([1]0)ab}_\ast V_{j}^{bc}U^{c j}
\label{expanast}
\end{eqnarray}
and 
\begin{eqnarray}
&&\hspace{-1mm}
A_i^{[0]}~=~A_i^{(0)}~=~{1\over g}(U_i+V_i)
\label{expani}\\
&&\hspace{-1mm}
A^{[1]}_i+A^{[2]}_i~=~A^{(1)}_i+A^{(2)}_i+O\big({m_\perp^7\over s^3}),~~~~~
A^{(1)}_i~=~{1\over p_\parallel^2}\tiL^{(0)}_i~~\sim~~{m_\perp^3\over s}
\nonumber\\
&&\hspace{-1mm}
A^{(2)a}_i~=~{1\over p_\parallel^2}\tiL^{(1)a}_i
+~{1\over p_\parallel^2}\big(\calp_\perp^2-\{\alpha,U_\bu\}-\{\beta,V_\ast\}\big)^{ab}A^{(1)b}_i
 -2i{1\over p_\parallel^2}(\calf^{[0]k}_i)^{ab}A^{(1)b}_k ~+~...
\nonumber
\end{eqnarray}
where ($n=1,2$)
\begin{equation}
\tiL^{(0)}_i~=~L^{(0)}_i+{4i\over s}\Big(V_{\bu i}{1\over p_\parallel^2}L^{(-1)}_\ast 
+U_{\ast i}{1\over p_\parallel^2}L^{(-1)}_\bu\Big) 
~=~L^{(0)}_i-{2i\over gs}(V_{\bu i}U^j)^{ab}{1\over\beta}V_j^b-{2i\over gs}(U_{\ast i}V^j)^{ab}{1\over\alpha}U_j^b
\label{til}
\end{equation}
In these formulas the singularity in ${1\over\alpha}$ is always causal ${1\over\alpha+i\epsilon}$ and similarly for ${1\over\beta}\equiv {1\over\beta+i\epsilon}$
and ${1\over p_\parallel^2}\equiv{1/s\over(\alpha+i\epsilon)(\beta+i\epsilon)}$, see Eq.  (\ref{propexplicit}).

The corresponding expansion of field strengths reads
\begin{eqnarray}
&&\hspace{-1mm}
gF^{(-1)a}_{\bu i}(x)~=~V^a_{\bu i}(x),~~~gF^{(-1)a}_{\ast i}(x)~=~U^a_{\ast i}(x)
\nonumber\\
&&\hspace{-1mm}
gF^{(0)a}_{\bu i}(x)~=~U^a_{\bu i}(x)-iU_\bu^{ab}(x)V_i^b(x)-{ig\over 2\alpha}\tiL_i^{(0)a}(x)+\cald_i^{ab}V_j^{bc}(x){1\over 2\alpha}U^{cj}(x)
\nonumber\\
&&\hspace{-1mm}
gF^{(0)a}_{\ast i}(x)~=~V^a_{\ast i}(x)-iV_\ast^{ab}(x)U^b_i(x)-{ig\over 2\beta}\tiL_i^{(0)a}(x)+\cald_i^{ab}U_j^{bc}(x){1\over 2\beta}V^{cj}(x)
\nonumber\\
&&\hspace{-1mm}
gF^{(-1)a}_{\ast\bu}(x)~=~U_{\ast\bu}^a(x)+V_{\ast\bu}^a(x)-{is\over 2}U_j^{ab}(x)V^{bj}(x)
\nonumber\\
&&\hspace{-1mm}
gF^{(0)a}_{ik}(x)~=~U_{ik}^a(x)+V_{ik}^a(x)-i\big(U_i^{ab}(x)V_k^b(x)-i\leftrightarrow k\big)
\label{Fields}
\end{eqnarray}
Power corrections to hadronic tensor  are proportional to
\begin{eqnarray}
&&\hspace{-5mm}
F^2(x)~\equiv~ F_{\mu\nu}^a(x)F^{a\mu\nu}(x)~=~{8\over s}F_{\bu i}^a(x)F_\ast^{ai}(x)+F_{ik}^a(x)F^{aik}(x)-{8\over s^2}F_{\ast\bu}^a(x)F_{\ast\bu}^a(x)
\nonumber\\
\label{fkva}
\end{eqnarray}
so
\begin{eqnarray}
&&\hspace{-1mm}
(F^2(x))^{(-1)}~=~ {8\over sg^2}U_{\ast i}^aV_\bu^{ai}
\nonumber\\
&&\hspace{-1mm}
(F^2(x))^{(0)}~=	~F^{(0)a}_{ik}(x)F^{(0)aik}-{8\over s^2}F^{(-1)a}_{\ast\bu}(x)F_{\ast\bu}^{(-1)a}(x)
\nonumber\\
&&\hspace{22mm}
+~{8\over sg}V_\bu^{ai}(x)F^{(0)a}_{\ast i}(x)
+{8\over sg}U_\ast^{ai}(x)F^{(0)a}_{\bu i}(x)
\label{fdva}
\end{eqnarray}
and the leading higher-twist correction is proportional to
\begin{eqnarray}
&&\hspace{-1mm}
(F^2(x))^{(0)}(F^2(0))^{(-1)}~+~(x\leftrightarrow 0)~=~\Big[F^{(0)a}_{ik}(x)F^{(0)aik}(x)
-{8\over s^2}F^{(-1)a}_{\ast\bu}(x)F_{\ast\bu}^{(-1)a}(x)
\nonumber\\
&&\hspace{-1mm}
+~{8\over sg}V_\bu^{ai}(x)F^{(0)a}_{\ast i}(x)
+{8\over sg}U_\ast^{ai}(x)F^{(0)a}_{\bu i}(x)\Big]{8\over sg^2}U_{\ast i}^a(0)V_\bu^{ai}(0)~+~(x\leftrightarrow 0)
\nonumber\\
\label{fkvadrat}
\end{eqnarray}
\section{Leading higher-twist correction at  $s\gg Q^2\gg Q_\perp^2\gg m^2$\label{sec:lhtc}}

As we mentioned in the Introduction, our method is relevant for calculation of higher-twist corrections
at any $s,Q^2\gg Q_\perp^2,m^2$. However, the expressions become manageable 
in the physically interesting case $s\gg Q^2\gg Q_\perp^2\gg m^2$ which we consider in this Section.
\footnote
{We also assume that the scalar particle is emitted in the central region of rapidity  so
$\alpha_qs\sim\beta_qs\gg Q^2$. \label{foot}}
We will demonstrate that the leading correction in this region comes from the following part of Eq. (\ref{fkva})
\begin{equation}
\hspace{-0mm}
g^2F^2(x)~=~{8\over s}U_\ast^{ai}(x)V_{\bu i}^a(x)
+2f^{mac}f^{mbd}\Delta^{ij,kl}U^a_i(x)U^b_j(x)V^c_k(x)V^d_l(x)~+~...
\label{fkvad}
\end{equation}
where
\begin{equation}
\hspace{-0mm}
\Delta^{ij,kl}~\equiv~g^{ij}g^{kl}-g^{ik}g^{jl}-g^{il}g^{jk}
\label{Delta}
\end{equation}
The higher-twist correction coming from the second term in the r.h.s. will be $\sim{Q_\perp^2\over Q^2}$
whereas other terms in the r.h.s. of Eq. (\ref{fkva}) yield contributions  $\sim{Q_\perp^2\over s}$,
$\sim{Q_\perp^2\over \alpha_qs}$, or $\sim{Q_\perp^2\over \beta_qs}$ all of which are small (see the footnote \ref{foot}).
In this approximation we get  
\begin{eqnarray}
&&\hspace{-1mm}
g^4F^2(x)F^2(0)~=~{64\over s^2}U_\ast^{mi}(x)V_{\bu i}^m(x)U_\ast^{nj}(0)V_{\bu j}^n(0)
\nonumber\\
&&\hspace{21mm}
+~{16\over s}f^{mac}f^{mbd}\Delta^{ij,kl}\big[U^a_i(x)U^b_j(x)V^c_k(x)V^d_l(x)
U_\ast^{nr}(0)V_{\bu r}^n(0)
\nonumber\\
&&\hspace{35mm}
+~U_\ast^{nr}(x)V_{\bu r}^n(x)
U^a_i(0)U^b_j(0)V^c_k(0)V^d_l(0)\big]
\label{mainff}
\end{eqnarray}
where the first term is the leading order and the second is the higher-twist correction.

Substituting our approximation (\ref{fkvad}) to Eq. (\ref{W}) and promoting background fields to operators as discussed in Sect. \ref{sec:funt} we get (note that $\alpha_q\beta_qs=Q_\parallel^2\simeq Q^2$):
\begin{eqnarray}
&&\hspace{-1mm}
W(p_A,p_B,q)~=~{64/s^2\over N_c^2-1}\!\int\! d^2x_\perp~
{2\over s}\!\int\!dx_\bu dx_\ast~\cos\big(\alpha_q x_\bu+\beta_qx_\ast-(q,x)_\perp\big)
\nonumber\\
&&\hspace{-1mm}
\times~\Big\{
\langle p_A|\hatU_\ast^{mi}(x_\bu,x_\perp)\hatU_\ast^{mj}(0)|p_A\rangle
\langle p_B|\hatV_{\bu i}^n(x_\ast,x_\perp)\hatV_{\bu j}^n(0)|p_B\rangle
\nonumber\\
&&\hspace{-1mm}
-~\frac{4N^2_c}{N^2_c - 4}~{\Delta^{ij,kl}\over Q^2}
~{2\over s}\!\int_{-\infty}^{x_\bu}\!dx'_\bu
~d^{abc}\langle p_A|\hatU^a_{\ast i}(x_\bu,x_\perp)\hatU^b_{\ast j}(x'_\bu,x_\perp)\hatU_{\ast r}^c(0)|p_A\rangle
\nonumber\\
&&\hspace{22mm}
\times~{2\over s}\!\int_{-\infty}^{x_\ast}\!dx'_\ast
~d^{mpq}\langle p_B|\hatV^m_{\bu k}(x_\ast,x_\perp)\hatV^p_{\bu l}(x'_\ast,x_\perp)\hatV_{\bu}^{qr}(0)|p_B\rangle\Big\}
\label{Wmain}
\end{eqnarray}
where we used formula \cite{Kaplan_Resnikoff_1967, MacFarlane:1968vc}
\begin{equation}
f^{acm}f^{bdm}d^{abn}d^{cdn}~=~\half(N_c^2-1)(N_c^2-4)
\label{formula}
\end{equation}

 Since an extra $U_{\ast k}$ (or $V_{\bu k}$) brings $s{x_i\over x_\perp^2}$
\footnote{To see this, we compared matrix elements of leading-twist operator 
$\langle p_A|U_\ast^{mi}(x_\bu,x_\perp)U_\ast^{mj}(0)|p_A\rangle$ and higher-twist
operator $\langle p_A|U^a_{\ast i}(x_\bu,x_\perp)U^b_{\ast j}(x'_\bu,x_\perp)U_{\ast r}^c(0)|p_A\rangle$ between quark states which gives an extra $s{x_r\over x_\perp^2}$ modulo some logarithms.
}
 we see that the higher-twist correction in the r.h.s of Eq. (\ref{Wmain}) is $\sim {q_\perp^2\over Q^2}$ so it gives the leading
 power  correction in the region $s\gg Q^2=m_\Phi^2\gg q_\perp^2 \gg m^2$. The TMD factorization formula 
 with  the higher-twist correction  (\ref{Wmain}) is the main result of the present paper.

We parametrize gluon TMD for unpolarized protons as (cf. Ref. \cite{Mulders:2000sh})
\begin{eqnarray}
&&\hspace{-1mm}
{4\over s^2g^2}\!\int\!dx_\ast\! \int\!d^2x_\perp~e^{-i\beta_q x_\ast+i(k,x)_\perp}
~\langle p_B|V^a_{\bu i}(x_\ast,x_\perp)V_{\bu j}^a(0)|p_B\rangle
\nonumber\\
&&\hspace{-1mm}
=~-\pi\beta_q\Big[g_{ij}D_g(\beta_q,k_\perp^2;\sigma_b)
-\big(2{k_i k_j\over m^2}+g_{ij}{k_\perp^2\over m^2}\big)
H_g(\beta_q,k_\perp^2;\sigma_b)\Big]
\label{parametrization}
\end{eqnarray}
where $\sigma_b$ is the cutoff in $\alpha$ integration in the target matrix elements,
see the discussion in Ref. \cite{Balitsky:2015qba}. The normalization here is such that 
$D_g(\beta_q,k_\perp^2;\sigma_b)$
is an unintegrated gluon distribution:
\begin{equation}
\int\! \dhd^2 k_\perp D_g(\beta_q,k_\perp^2;\sigma_b)~=~D_g(\beta_q,\mu^2=\sigma_b\beta_qs)
\end{equation}
where
$D_g(\beta_q,\mu^2)$ is the usual gluon parton density (this formula is correct in the leading log approximation, see
the discussion in Ref. \cite{Balitsky:2015qba}).

Next,  the three-gluon matrix element  in Eq. (\ref{Wmain}) for unpolarized hadrons can  be
parametrized as
\begin{eqnarray}
&&\hspace{-1mm}
{4\over s^2g^3}\!\int\!dx_\ast~\! \int\!d^2x_\perp~e^{-i\beta_q x_\ast+i(k,x)_\perp}
\!\int_{-\infty}^{x_\ast}\!d{2\over s}x'_\ast
~d^{abc}\langle p_B|V^a_{\bu i}(x_\ast,x_\perp)V^b_{\bu j}(x'_\ast,x_\perp)
V_{\bu r}^c(0)|p_B\rangle~+~i\leftrightarrow j
\nonumber\\
&&\hspace{-1mm}
=~-\pi\beta_q\Big[(k_ig_{jr}+k_jg_{ir})D_1^g(\beta_q,k_\perp^2;\sigma_b)
+k_rg_{ij}D_2^g(\beta_q,k_\perp^2;\sigma_b)
\nonumber\\
&&\hspace{-1mm}-\big[k_i k_j k_r + {k_\perp^2\over 4}(k_rg_{ij}+k_ig_{jr}+k_jg_{ir})\big]
{1\over m^2}H_1^g(\beta_q,k_\perp^2;\sigma_b)\Big]
\label{parametriz}
\end{eqnarray}
At large $k_\perp^2$ gluon TMDs in the r.h.s. of Eq. (\ref{parametrization}) behave as 
$D_g(\beta_q, k^2_\perp)\sim{1\over k_\perp^2}$ 
and $H_g(\beta_q,k_\perp^2)\sim{1\over k_\perp^4}$. Similarly, one should expect that $D_i^g(\beta_q,k^2_\perp)\sim{1\over k_\perp^2}$ 
and $H_1^g(\beta_q,k_\perp^2)\sim{1\over k_\perp^4}$.

It is well known that in our kinematic region $s\gg Q^2\gg Q_\perp^2$ 
gluon TMDs (\ref{parametrization}) possess Sudakov logs of the type 
\begin{equation}
{4\over s^2g^2}\int\! dx_\ast \int\!d^2x_\perp~e^{-i\beta_q x_\ast+i(k,x)_\perp}
\langle p_B|V_{\bu i}^n(x_\ast,x_\perp)V_\bu^{ni}(0)|p_B\rangle
~\sim~e^{-{\alpha_sN_c\over 2\pi}
\ln^2{\sigma_b s\over k_\perp^2}} D_g(\beta_q,k_\perp,\ln{k_\perp^2\over s})
\label{sudakov}
\end{equation}
One should expect double-logs of this type in 
$D_i^g(\beta_q,k_\perp^2;\sigma_b)$ and  $H_1^g(\beta_q,k_\perp^2;\sigma_b)$, too.

Let us now demonstrate that the terms in $(F^2(x))^{(0)}$ (see Eq. (\ref{fdva})) which we neglected give  small 
contributions.
For example, consider the following contribution to $F^2(x)F^2(0)$:
\begin{eqnarray}
&&\hspace{-1mm}
-{64i\over s^2}U_\ast^{ai}(x)V_{\bu i}^a(x)V_\bu^{bj}(0)V_\ast^{bc}(0)U^c_j(0)
\end{eqnarray}
The corresponding contribution to hadronic tensor $W$ has the form
\begin{eqnarray}
&&\hspace{-1mm}
-{64/s^2\over N_c^2-1}\!\int\! d^2x_\perp~e^{i(q,x)_\perp}
{2\over s}\!\int\!dx_\bu dx_\ast~e^{-i\alpha_q x_\bu-i\beta_qx_\ast}
\nonumber\\
&&\hspace{-1mm}
\times~{2\over\alpha_q s}\langle p_A|U^{ai}_{\ast}(x_\bu,x_\perp)U^{aj}_{\ast}(0)|p_A\rangle
\langle p_B|V^b_{\bu i}(x_\ast,x_\perp)V^{bc}_{\ast}(0)V_{\bu j}^c(0)|p_B\rangle
\label{Wexample1}
\end{eqnarray}
Note that unlike Eq. (\ref{Wmain}), the factor in the denominator is $\alpha_qs\gg Q^2$ so the 
contribution (\ref{Wexample1}) is power suppressed in comparison to   Eq. (\ref{Wmain}) in our kinematic region.
\footnote{Of course, this power suppression may be moderated by difference in logarithmic evolution 
of operators in the r.h.s.'s of Eqs. (\ref{Wmain}) and (\ref{Wexample1}), but one should expect the evolution
of these operators to be of the same order of magnitude.}

As a less trivial example, consider the  following term in $F^2(x)F^2(0)$
\begin{eqnarray}
&&\hspace{-1mm}
-{64\over s^3}U_{\ast i}^a(x)V_\bu^{ai}(x)
V_\bu^{bj}(0){1\over \beta}(V_{\bu j}U^k)^{bc}{1\over\beta}V_k^c(0)
\label{xz1}
\end{eqnarray}
The corresponding contribution to hadronic tensor $W$ reads
\begin{eqnarray}
&&\hspace{-1mm}
{64/s^2\over N_c^2-1}\!\int\! d^2x_\perp~e^{i(q,x)_\perp}
{2\over s}\!\int\!dx_\bu dx_\ast~e^{-i\alpha_q x_\bu-i\beta_qx_\ast}
\Big\{
{i\over\alpha_q s}\langle p_A|U_\ast^{mi}(x_\bu,x_\perp)U_\ast^{mj}(0)|p_A\rangle
\nonumber\\
&&\hspace{-1mm}
\times~
{4\over s^2}\!\int_{-\infty}^{0_\ast} dz_\ast \!\int _{-\infty}^{z_\ast}\!dz'_\ast~ (z-z')_\ast
\langle p_B|V_{\bu i}^a(x_\ast,x_\perp)V_\bu^{bk}(0)
(V_{\bu k}(z_\ast,0_\perp)T^a)^{bc}V_{\bu j}^c(z'_\ast,0_\perp)|p_B\rangle\Big\}
\nonumber\\
\label{Wexample2}
\end{eqnarray}
where we used
$$
{1\over\beta+i\epsilon}V_k(x)~=~-i\int_{-\infty}^{x_\ast}dx'_\ast~V_k(x'_\ast, x_\perp)
~=~-{2i\over s}\int_{-\infty}^{x_\ast}dx'_\ast~(x-x')_\ast V_{\bu k}(x'_\ast, x_\perp)
$$

In both examples (\ref{Wexample1}) and (\ref{Wexample2})  the factor ${1\over \alpha_q}$  comes from an extra integration 
over $x'_\bu$  in $U_i$, see  Eq. (\ref{uiv}):
\begin{eqnarray}
&&\hspace{-1mm}
\int\! dx_\bu ~e^{-i\alpha_q x_\bu}\langle U_i(x_\bu,x_\perp)U_j(0)\rangle~=~
{2\over s}\!\int\! dx_\bu\!\int_{-\infty}^{x_\bu}\! dx'_\bu e^{-i\alpha_q x_\bu}\langle U_{\ast i}(x'_\bu,x_\perp)U_j(0)\rangle
\nonumber\\
&&\hspace{-1mm}
=~
-{2i\over \alpha_q s}\!\int\! dx_\bu~e^{-i\alpha x_\bu}\langle U_{\ast i}(x_\bu,x_\perp)U_j(0)\rangle
\label{xz}
\end{eqnarray}

 The way to figure out such integrations
is very simple: take $\alpha_q\rightarrow 0$ and check if there is an infinite
integration of the type $\int_{-\infty}^{x_\bu}dx'_\bu$. Evidently, it may happen if 
we have a single $U_i(x)$ (without any additional $U$-operators) at the point $x$, or a single $U_i(0)$.

Similarly, the factor ${1\over\beta_q}$ 
comes from  an extra integration over $x'_\ast$ in $V_i$ in  Eq. (\ref{uiv})  so an indication of such 
contribution is the infinite integration $\int_{-\infty}^{x_\ast}dx'_\ast$ in the limit $\beta_q\rightarrow 0$
which translates to the condition of a single $V_i$ at the point $x$ or at the point $0$.

Thus, to get the terms $\sim{1\over Q^2}$ we need to find contributions which satisfy 
both of the above conditions which singles out the contribution (\ref{mainff}).

\section{Small-x limit and scattering of shock waves\label{sec:sx}}

Let us consider the hadronic tensor 
\begin{equation}
\langle p_A,p_B|g^4F^2(x)F^2(y) |p_A,p_B\rangle 
\end{equation}
in the small-x limit $s\rightarrow\infty$, $Q^2$ and $q_\perp^2$ - fixed. At first, let us not impose the condition $Q^2\gg q_\perp^2$ 
which means that the relation between $x_\parallel^2$ and $x_\perp^2$ is arbitrary
(later we will see that  $Q^2\gg q_\perp^2$ corresponds to $x_\parallel^2\ll x_\perp^{2}$). 

The small-x limit may be obtained by rescaling $s\rightarrow\lambda^2 s\Leftrightarrow p_1\rightarrow\lambda p_1,p_2\rightarrow\lambda p_2$.
As discussed in Refs. \cite{Balitsky:1995ub,Balitsky:1998ya,Balitsky:2004rr}, the only components of field strength surviving in this rescaling are 
$U_{\ast i}(x_\bu,x_\perp)$ and $V_{\bu i}(x_\ast,x_\perp)$. 
Moreover, if we study classical fields at longitudinal distances which does not scale with $\lambda$,  
we  can replace the projectile and target fields by 
infinitely thin ``shock waves''
\begin{equation}
U_{\ast i}(x_\bu,x_\perp)~\rightarrow~{s\over 2}\delta(x_\bu)\calu_i(x_\perp)~~~~~{\rm and}~~~~
V_{\bu i}(x_\ast,x_\perp)~\rightarrow~{s\over 2}\delta(x_\ast)\calv_i(x_\perp)
\label{shockwave}
\end{equation}
However, since we need to compare the classical fields in the small-$x$ limit to 
 our expressions (\ref{Fields}) at small 
longitudinal distances, we will keep $x_\ast$ and $x_\bu$ dependence for a while.

As described above, to find the classical fields we can 
start with the trial configuration
\begin{eqnarray}
&&\hspace{-1mm}
g\cala_i^{[0]}(x)~=~U_i(x_\bu,x_\perp)+V_i(x_\ast,x_\perp),~~~~~\cala_\ast^{[0]}=\cala_\bu^{[0]}~=~0
\nonumber\\
&&\hspace{-1mm}
\Psi^{[0]}(x)~=~\Sigma_a(x_\bu,x_\perp)+\Sigma_b(x_\ast,x_\perp),~~~~~
\notp_1\Sigma_a=\notp_2\Sigma_b=\gamma_i\Sigma_a=\gamma_i\Sigma_b~=~0
\label{trialsx}
\end{eqnarray}
with the linear term
\begin{eqnarray}
&&\hspace{-1mm}
gL_\mu^a~=~{2ip_{1\mu}\over s}V^{jab}U_{\ast j}^b+{2ip_{2\mu}\over s}U^{jab}V_{\bu j}^b
-i\cald_j^{ab}(U^{jbc}V_\mu^{\perp c}+V^{jbc}U_\mu^{\perp c})
\label{lintermx}
\end{eqnarray}
and improve it order by order in $L_\mu$. In this way we'll get a set of Feynman diagrams in the background field
(\ref{trialsx}). Unfortunately, in the general case of arbitrary relation between $q_\parallel$ and $q_\perp$ we no
longer have a small parameter ${p_\perp^2\over p_\parallel^2}$ so we need explicit expressions for propagators in the background fields, and, in addition, we need all orders in the expansion of linear term (\ref{lintermx}).  
Still, we can compare our calculations with the perturbative expansion of classical fields in powers of the ``parameter'' 
$[U_i,V_j]$ carried out in Refs. \cite{Balitsky:1998ya, Balitsky:2005we}. In the leading order in perturbation theory only the first line of Eq. (\ref{fields}) survives and we get
\begin{eqnarray}
&&\hspace{-1mm}
gA_\bu~=~{i\over p^2+i\epsilon p_0}[U^j,V_{\bu j}],~~~~gA_\ast~=~{i\over p^2+i\epsilon p_0}[V^j,U_{\ast j}]~~~~~
\nonumber\\
&&\hspace{-1mm}
gA_i~=~U_i+V_i+{p^j\over p^2+i\epsilon p_0}\big([U_i,V_j]-i\leftrightarrow j\big)
\end{eqnarray}
The corresponding expressions for field strengths are
\begin{eqnarray}
&&\hspace{-2mm}
gF_{\bu i}~=~V_{\bu i}-{p^j\over p^2+i\epsilon p_0}\big(g_{ij}[U^k,V_{\bu k}]+[U_j,V_{\bu i}]-[U_i,V_{\bu j}]\big)~~~~~
\label{filds}\\
&&\hspace{-2mm}
gF_{\ast i}~=~U_{\ast i}-{p^j\over p^2+i\epsilon p_0}\big(g_{ij}[V^k,U_{\ast k}]+[V_j,U_{\ast i}]-[V_i,U_{\ast j}]\big)
\nonumber\\
&&\hspace{-2mm}
gF_{\ast\bu}~=~{2i\over p^2+i\epsilon p_0}[U_\ast^{~j},V_{\bu j}]~~~~
\nonumber\\
&&\hspace{-2mm}
gF_{ij}~=-i[U_i,V_j]-{ip_ip^k\over p^2+i\epsilon p_0}([U_j,V_k]-j\leftrightarrow k)-i\leftrightarrow j
~=~{4i/s\over p^2+i\epsilon p_0}([U_{\ast i},V_{\bu j}]-i\leftrightarrow j)
\nonumber
\end{eqnarray}
In the last line we used the identity
\begin{equation}
p_i([U_j,V_k]-j\leftrightarrow k)-i\leftrightarrow j~=~-p_k([U_i,V_j]-i\leftrightarrow j)
\end{equation}
and the fact that in the small-$x$ limit $\partial_iU_j-\partial_jU_i-i[U_i,U_j]~=~\partial_iV_j-\partial_jV_i-i[V_i,V_j]~=~0$.

Let us discuss now how our approximation ${p_\perp^2\over p_\parallel^2}\ll 1$ looks in the coordinate space. 
The explicit expressions for fields (\ref{filds}) are
\begin{eqnarray}
&&\hspace{-1mm}
gF_{\bu i}(x)~=~V_{\bu i}(x_\ast,x_\perp)
+~{i\over 4\pi}\!\int\! dz~{1\over (x-z)_\ast}{\partial\over\partial x_j}\theta\big[(x-z)_\parallel^2-(x-z)_\perp^2\big]
\theta(x-z)_\ast gL^-_{ij}(z)
\nonumber\\
&&\hspace{-1mm}
gF_{\ast i}(x)~=~U_{\ast i}(x_\bu,x_\perp)
-~{i\over 4\pi}\!\int\! dz~{1\over (x-z)_\bu}{\partial\over\partial x_j}\theta\big[(x-z)_\parallel^2-(x-z)_\perp^2\big]
\theta(x-z)_\bu gL^+_{ij}(z)
\nonumber\\
&&\hspace{-1mm}
gF_{\ast\bu}(x)
~=~-{i\over \pi}\!\int\! dz~\delta\big[(x-z)_\parallel^2-(x-z)_\perp^2\big]
\theta(x-z)_\ast [U_\ast^{~j}(z_\bu,z_\perp),V_{\bu j}(z_\ast,z_\perp)]
\label{fiildz}\\
&&\hspace{-1mm}
gF_{ij}(x)
~=~-{2i\over \pi s}\!\int\! dz~\delta\big[(x-z)_\parallel^2-(x-z)_\perp^2\big]
\theta(x-z)_\ast \big([U_{\ast i}(z_\bu,z_\perp),V_{\bu j}(z_\ast,z_\perp)]-i\leftrightarrow j\big)
\nonumber
\end{eqnarray}
where 
\begin{equation}
gL^{\pm}_{ij}(z)~\equiv~g_{ij}[U_\ast^{~k},V_{\bu k}]\pm [U_{\ast i},V_{\bu j}]\mp[U_{\ast j},V_{\bu i}]
\end{equation}
At longitudinal distances $x_\bu,x_\ast\sim 1$ these expressions agree with Eq. (52) from Ref. \cite{Balitsky:1998ya} 
after the replacement (\ref{shockwave}).

Now let us compare the fields (\ref{fiildz}) at small longitudinal distances to our approximate solution (\ref{Fields}). 
Let us start with $F_{ij}(x)$ in the last line in Eq. (\ref{fiildz}). If $(x-z)_\parallel^2$ is smaller than the characteristic 
transverse distances in the integral over $z_\perp$ one can replace $[U_{\ast i}(z_\bu,z_\perp),V_{\bu j}(z_\ast,z_\perp)]$
by $[U_{\ast i}(z_\bu,x_\perp),V_{\bu j}(z_\ast,x_\perp)]$  and get
\begin{eqnarray}
&&\hspace{-1mm}
gF_{ij}(x)
~=~-{2i\over s}\!\int\! d^2z_\parallel
\theta(x-z)_\ast\theta(x-z)_\bu \big([U_{\ast i}(z_\bu,x_\perp),V_{\bu j}(z_\ast,x_\perp)] -i\leftrightarrow j\big)
\nonumber\\
&&\hspace{-1mm}~=~-i[U_i(x_\bu,x_\perp),V_j(x_\ast,x_\perp)]+i[U_j(x_\bu,x_\perp),V_i(x_\ast,x_\perp)]
\end{eqnarray}
which is exactly the last line in Eq. (\ref{Fields}). Similarly, the third line in Eq. (\ref{fiildz}) reproduces $F_{\ast\bu}$ in the
fourth line in Eq. (\ref{Fields}). 

Next, $gF^{(0)a}_{\bu i}$ in  second line  in Eq. (\ref{Fields}) in the leading order in perturbation theory turns to
\begin{equation}
-{\partial^j\over 2\alpha}(g_{ij}[U^k,V_k]-[U_i,V_j]+[U_j,V_i])
~=~{2i\over s^2}\!\int_{-\infty}^{x_\ast}\! dz_\ast \!\int_{-\infty}^{x_\bu}\! dz_\bu 
~(x-z)_\bu \partial^jL^-_{ij}(z_\ast,z_\bu,x_\perp)
\label{5.11}
\end{equation}
On the other hand,  the first line in Eq. (\ref{fiildz}) at small $(x-z)_\parallel$ gives
\begin{eqnarray}
&&\hspace{-1mm}
{i\over 4\pi}\!\int\! dz~{\theta(x-z)_\ast\over (x-z)_\ast}\theta\big[(x-z)_\parallel^2-(x-z)_\perp^2\big]
 {\partial\over\partial z_j}L^-_{ij}(z)
\nonumber\\
 &&\hspace{-1mm}
 \simeq~
 {i\over 4\pi}\!\int\! dz~{\theta(x-z)_\ast\over (x-z)_\ast}\theta\big[{4\over s}(x-z)_\ast(x-z)_\bu-(x-z)_\perp^2\big]
\partial^j L^-_{ij}(z_\ast,z_\bu,x_\perp)
\end{eqnarray}
which agrees with Eq. (\ref{5.11}) after integration over $z_\perp$. Similarly, one can check the consistency of
two expressions for $F_{\ast i}$.

\section{Conclusions and outlook}
We have formulated the approach to TMD factorization based on the factorization in rapidity
and found the leading higher-twist contribution to the production of a scalar particle (e.g. Higgs) 
by gluon-gluon fusion in the hadron-hadron scattering. 
Up to now our results are obtained in the tree-level approximation when the question of exact matching of cutoffs in rapidity does not arise. However, this question will 
become crucial starting from  the
first loop. In our previous papers we calculated the evolution of gluon TMD with respect to our
rapidity cutoff so we need to match it to the coefficient functions in front of TMD operators. 
The work is in progress.

Also, we obtained power corrections for particle production only in the case of gluon-gluon fusion. 
It would be interesting (and we plan) to find power corrections to Drell-Yan process. 
There is a statement that for semi-inclusive deep inelastic scattering (SIDIS) 
the leading-order TMDs have different directions of Wilson lines: one to $+\infty$ and
another to $-\infty$. We think that the same directions of Wilson lines will be 
in the case of power corrections and we plan to study this question in forthcoming publications.

The authors are grateful to J.C. Collins, S. Dawson, A. Kovner, D. Neill, A. Prokudin, T. Rogers, 
and R. Venugopalan for valuable discussions. This material is based upon work supported by the U.S. Department of Energy, Office of Science, Office of Nuclear Physics under contracts DE-AC02-98CH10886 and DE-AC05-06OR23177.

\section{Appendix A}
In this Section we prove that  the field $C_\mu$ created by a source $J_\mu$ 
in the presence of external fields $\barA_\mu$ and $\barB_\mu$  
\footnote{For simplicity, in this section we disregard quarks so in our case $J_\mu$ is Eq. (\ref{linterm}) without quark terms.}
\begin{eqnarray}
&&\hspace{-1mm}
\langle C^a_\mu(x)\rangle_J~\equiv~\int\! D\tilC DC~C^a_\mu(x)~
\exp\Big\{\!\int\! dz~\Big[{i\over 2}\tilC^{m\xi}\Box^{mn}_{\xi\eta}\tilC^{n\eta}
\nonumber\\
&&\hspace{-1mm}
+~igf^{mnl}\barD^\xi \tilC^{m\eta} \tilC^n_\xi \tilC^l_\eta 
+\frac{ig^2}{4}f^{abm}f^{cdm}\tilC^{a\xi} \tilC^{b\eta}\tilC^c_\xi \tilC^d_\eta
-iJ^m_\xi\tilC^{m\xi}-{i\over 2}C^{m\xi}\Box^{mn}_{\xi\eta}C^{n\eta}
\nonumber\\
&&\hspace{-1mm}
-~igf^{mnl}\barD^\xi C^{m\eta} C^n_\xi C^l_\eta 
-{ig^2\over 4}f^{abm}f^{cdm}C^{a\xi} C^{b\eta}C^c_\xi C^d_\eta+iJ^m_\xi C^{m\xi}\Big]\Big\}
\label{a1}
\end{eqnarray}
 is given by a set of Feynman
diagrams with retarded Green functions  (note that Eq. (\ref{a1})  implies that  $J_\mu, \barA_\mu$, and  $\barB_\mu$ are the same in the right and left part of the amplitude). Hereafter
we use the notation $\Box_{\mu\nu}\equiv \barP^2 g_{\mu\nu}+2i\barG_{\mu\nu}$.

First, we consider gluon propagators for the double functional integral  over $C$ fields in the background 
filelds $\barA=\bar\tilA$,  $\barB=\bar\tilB$
and prove that
\begin{eqnarray}
&&\hspace{-1mm}
\langle C^a_\mu(x)C^b_\nu(y)\rangle - \langle C^a_\mu(x)\tilC^b_\nu(y)\rangle
~=~(x|{-i\over \Box^{\mu\nu}+i\epsilon p_0}|y)^{ab}
\nonumber\\
&&\hspace{-1mm}
\langle \tilC^a_\mu(x)C^b_\nu(y)\rangle - \langle \tilC^a_\mu(x)\tilC^b_\nu(y)\rangle
~=~(x|{-i\over \Box^{\mu\nu}+i\epsilon p_0}|y)^{ab}
\label{a2}
\end{eqnarray}
Note that  we define $\langle \mato\rangle$ in this Section as
\begin{equation}
\langle \mato\rangle~\equiv~\int\! D\tilC DC~\mato~
e^{\int\! dz\big({i\over 2}\tilC^{a\mu}\Box^{ab}_{\mu\nu}\tilC^{b\nu}
-{i\over 2}C^{a\mu}\Box^{ab}_{\mu\nu}C^{b\nu}\big)}
\label{dabel}
\end{equation}

To prove Eq. (\ref{a2}), we write down 
\begin{equation}
\Box_{\mu\nu}~=~p^2g_{\mu\nu}+\calo_{\mu\nu},~~~~
\calo_{\mu\nu}~\equiv~\big(\{p^\xi,\barA_\xi+\barB_\xi\}+(\barA+\barB)^2\big)g_{\mu\nu}+2i\barG_{\mu\nu}
\end{equation}
and expand in powers of $\calo_{\mu\nu}$.

In the trivial order Eqs. (\ref{a2}) immediately follow from the bare propagators
for the double functional integral (\ref{dabel})
\begin{eqnarray}
&&\hspace{-1mm}
\langle C^a_\mu(x)C^b_\nu(y)\rangle_{\rm bare}~=~(x|{-ig_{\mu\nu}\delta^{ab}\over p^2+i\epsilon}|y),~~~~~~
\langle  \tilC^a_\mu(x)\tilC^b_\nu(y)\rangle_{\rm bare}~=~(x|{ig_{\mu\nu}\delta^{ab}\over p^2-i\epsilon}|y)
\nonumber\\
&&\hspace{-1mm}
\langle  C^a_\mu(x)\tilC^b_\nu(y)\rangle_{\rm bare}~=~-g_{\mu\nu}\delta^{ab}(x|2\pi\delta(p^2)\theta(-p_0)|y)
\label{bareprops}
\end{eqnarray}
where 
\begin{equation}
\langle \mato\rangle_{\rm bare}~\equiv~\int\! D\tilC DC~\mato~
e^{\int\! dz\big({i\over 2}C^{a\mu}\partial^2C^a_\mu
-{i\over 2}\tilC^{a\mu}\partial^2\tilC^a_\mu\big)}
\label{dabare}
\end{equation}

In the first order in $\calo_{\mu\nu}$ we get
\begin{eqnarray}
\hspace{-1mm}
\langle C^a_\mu(x)C^b_\nu(y)\rangle^{(1)}
&~=~&i\!\int\! dz\big[-
\langle C^a_\mu (x)C^{c\xi}(z)\rangle_{\rm bare}\calo_{\xi\eta}^{cd}(z)
\langle C^{d\eta}(z) C^b_\nu(y)\rangle_{\rm bare}
\nonumber\\
&+&\langle C^a_\mu (x)\tilC^{c\xi}(z)\rangle_{\rm bare}
\calo_{\xi\eta}^{cd}(z)\langle \tilC^{d\eta}(z) C^b_\nu(y)\rangle_{\rm bare}\big]
\nonumber\\
\langle C^a_\mu(x)\tilC^b_\nu(y)\rangle^{(1)}
&~=~&i\!\int\! dz\big[-
\langle C^a_\mu (x)C^{c\xi}(z)\rangle_{\rm bare}\calo_{\xi\eta}^{cd}(z)
\langle C^{d\eta}(z) \tilC^b_\nu(y)\rangle_{\rm bare}
\nonumber\\
&+&\langle C^a_\mu (x)\tilC^{c\xi}(z)\rangle_{\rm bare}
\calo_{\xi\eta}^{cd}(z)\langle \tilC^{d\eta}(z) \tilC^b_\nu(y)\rangle_{\rm bare}\big]
\label{a8}
\end{eqnarray}
so
\begin{eqnarray}
&&\hspace{-1mm}
\langle C^a_\mu(x)C^b_\nu(y)\rangle^{(1)} - \langle C^a_\mu(x)\tilC^b_\nu(y)\rangle^{(1)}
~=~i(x|{1\over p^2+i\epsilon p_0}\calo_{\mu\nu}^{ab}{1\over p^2+i\epsilon p_0}|y)
\label{a9}
\end{eqnarray}
Similarly, it is easy to see that 
\begin{eqnarray}
\hspace{-1mm}
\langle \tilC^a_\mu(x)C^b_\nu(y)\rangle^{(1)}
&~=~&i\!\int\! dz\big[-
\langle \tilC^a_\mu (x)C^{c\xi}(z)\rangle_{\rm bare}\calo_{\xi\eta}^{cd}(z)
\langle C^{d\eta}(z) C^b_\nu(y)\rangle_{\rm bare}
\nonumber\\
&+&\langle \tilC^a_\mu (x)\tilC^{c\xi}(z)\rangle_{\rm bare}
\calo_{\xi\eta}^{cd}(z)\langle \tilC^{d\eta}(z) C^b_\nu(y)\rangle_{\rm bare}\big]
\nonumber\\
\langle \tilC^a_\mu(x)\tilC^b_\nu(y)\rangle^{(1)}
&~=~&i\!\int\! dz\big[-
\langle \tilC^a_\mu (x)C^{c\xi}(z)\rangle_{\rm bare}\calo_{\xi\eta}^{cd}(z)
\langle C^{d\eta}(z) \tilC^b_\nu(y)\rangle_{\rm bare}
\nonumber\\
&+&\langle \tilC^a_\mu (x)\tilC^{c\xi}(z)\rangle_{\rm bare}
\calo_{\xi\eta}^{cd}(z)\langle \tilC^{d\eta}(z) \tilC^b_\nu(y)\rangle_{\rm bare}\big]
\label{a10}
\end{eqnarray}
so
\begin{eqnarray}
&&\hspace{-1mm}
\langle \tilC^a_\mu(x)C^b_\nu(y)\rangle^{(1)} - \langle \tilC^a_\mu(x)\tilC^b_\nu(y)\rangle^{(1)}
~=~i(x|{1\over p^2+i\epsilon p_0}\calo_{\mu\nu}^{ab}{1\over p^2+i\epsilon p_0}|y)
\label{a11}
\end{eqnarray}

In the second order in $\calo_{\mu\nu}$ 
\begin{eqnarray}
\hspace{-1mm}
\langle C^a_\mu(x)C^b_\nu(y)\rangle^{(2)}
&~=~&i\!\int\! dz\big[-
\langle C^a_\mu (x)C^{c\xi}(z)\rangle^{(1)}\calo_{\xi\eta}^{cd}(z)
\langle C^{d\eta}(z) C^b_\nu(y)\rangle_{\rm bare}
\nonumber\\
&+&\langle C^a_\mu (x)\tilC^{c\xi}(z)\rangle^{(1)}
\calo_{\xi\eta}^{cd}(z)\langle \tilC^{d\eta}(z) C^b_\nu(y)\rangle_{\rm bare}\big]
\nonumber\\
\langle C^a_\mu(x)\tilC^b_\nu(y)\rangle^{(2)}
&~=~&i\!\int\! dz\big[-
\langle C^a_\mu (x)C^{c\xi}(z)\rangle^{(1)}\calo_{\xi\eta}^{cd}(z)
\langle C^{d\eta}(z) \tilC^b_\nu(y)\rangle_{\rm bare}
\nonumber\\
&+&\langle C^a_\mu (x)\tilC^{c\xi}(z)\rangle^{(1)}
\calo_{\xi\eta}^{cd}(z)\langle \tilC^{d\eta}(z) \tilC^b_\nu(y)\rangle_{\rm bare}\big]
\label{a12}
\end{eqnarray}
so using the results (\ref{a9}) and (\ref{a11}) we get
\begin{equation}
\hspace{-1mm}
\langle C^a_\mu(x)C^b_\nu(y)\rangle^{(2)} - \langle C^a_\mu(x)\tilC^b_\nu(y)\rangle^{(2)}
~=~-i(x|{1\over p^2+i\epsilon p_0}\calo_{\mu\xi}{1\over p^2+i\epsilon p_0}
\calo^\xi_{~\nu}{1\over p^2+i\epsilon p_0}|y)^{ab}
\label{a13}
\end{equation}
Similarly, it is  easy to demonstrate that 
\begin{equation}
\hspace{-1mm}
\langle \tilC^a_\mu(x)C^b_\nu(y)\rangle^{(2)} - \langle \tilC^a_\mu(x)\tilC^b_\nu(y)\rangle^{(2)}
~=~-i(x|{1\over p^2+i\epsilon p_0}\calo_{\mu\xi}{1\over p^2+i\epsilon p_0}
\calo^\xi_{~\nu}{1\over p^2+i\epsilon p_0}|y)^{ab}
\label{a14}
\end{equation}
One can prove now Eq. (\ref{a2}) by induction using formulas
\begin{eqnarray}
\hspace{-1mm}
\langle C^a_\mu(x)C^b_\nu(y)\rangle^{(n)}
&~=~&i\!\int\! dz\big[-
\langle C^a_\mu (x)C^{c\xi}(z)\rangle^{(n-1)}\calo_{\xi\eta}^{cd}(z)
\langle C^{d\eta}(z) C^b_\nu(y)\rangle_{\rm bare}
\nonumber\\
&+&\langle C^a_\mu (x)\tilC^{c\xi}(z)\rangle^{(n-1)}
\calo_{\xi\eta}^{cd}(z)\langle \tilC^{d\eta}(z) C^b_\nu(y)\rangle_{\rm bare}\big]
\nonumber\\
\langle C^a_\mu(x)\tilC^b_\nu(y)\rangle^{(n)}
&~=~&i\!\int\! dz\big[-
\langle C^a_\mu (x)C^{c\xi}(z)\rangle^{(n-1)}\calo_{\xi\eta}^{cd}(z)
\langle C^{d\eta}(z) \tilC^b_\nu(y)\rangle_{\rm bare}
\nonumber\\
&+&\langle C^a_\mu (x)\tilC^{c\xi}(z)\rangle^{(n-1)}
\calo_{\xi\eta}^{cd}(z)\langle \tilC^{d\eta}(z) \tilC^b_\nu(y)\rangle_{\rm bare}\big]
\label{a15}
\end{eqnarray}

Now we are in a position to prove Eq. (\ref{a1}).
In the leading order in $g$ it is trivial: using Eqs. (\ref{a2}) one immediately sees that
\begin{eqnarray}
&&\hspace{-1mm}
\langle C^a_\mu(x)\rangle^{[0]}_J~=~\int\! D\tilC DC~C^a_\mu(x)~
e^{\int\! dz\big({i\over 2}\tilC^{a\xi}\Box^{ab}_{\xi\eta}\tilC^{b\eta}-iJ^a_\xi\tilC^{a\xi}
-{i\over 2}C^{a\xi}\Box^{ab}_{\xi\eta}C^{b\eta}+iJ^a_\xi C^{a\xi}\big)}
\nonumber\\
&&\hspace{-1mm}
=~\int\! D\tilC DC~\tilC_\mu(x)~
e^{\int\! dz\big(-{i\over 2}\tilC^{a\xi}\Box^{ab}_{\xi\eta}\tilC^{b\eta}-iJ^a_\xi\tilC^{a\xi}
+{i\over 2}C^{a\xi}\Box^{ab}_{\xi\eta}C^{b\eta}+iJ^a_\xi C^{a\xi}\big)}
\nonumber\\
&&\hspace{-1mm}
=~\int\! dz~(x|{1\over \Box^{\mu\nu}+i\epsilon p_0}|z)^{ab}J^{b\nu}(z)
\label{a16}
\end{eqnarray}

In the first order in $g$ (with one three-gluon vertex) we obtain
\begin{eqnarray}
&&\hspace{-1mm}
\langle C^a_\mu(x)\rangle_J^{[1]}~=~-igf^{mnl}\!\int\! D\tilC DC~C^a_\mu(x)\!\int\! dz 
\big[\barD^\xi C^{m\eta} C^n_\xi C^l_\eta(z)-\barD^\xi \tilC^{m\eta} \tilC^n_\xi \tilC^l_\eta(z)\big]
\nonumber\\
&&\hspace{22mm}
\times~\exp\Big\{\!\int\! dz'~\big[{i\over 2}\tilC^{a\xi}\Box^{ab}_{\xi\eta}\tilC^{b\eta}
-iJ^a_\xi\tilC^{a\xi}
-{i\over 2}C^{a\xi}\Box^{ab}_{\xi\eta}C^{b\eta}+iJ^a_\xi C^{a\xi}\big](z')\Big\}
\nonumber\\
&&\hspace{-1mm}
=~\frac{ig}{2}f^{mnl}\!\int\! dz dz'dz''\langle C^a_\mu(x)
\big[\barD^\xi C^{m\eta} C^n_\xi C^l_\eta(z)-\barD^\xi \tilC^{m\eta} \tilC^n_\xi \tilC^l_\eta(z)\big]
\nonumber\\
&&\hspace{22mm}
\times~[J^c_\alpha C^{c\alpha}(z')-J^c_\alpha \tilC^{c\alpha}(z')]
[J^d_\beta C^{d\beta}(z'')-J^d_\beta \tilC^{d\beta}(z'')
\rangle
\nonumber\\
&&\hspace{-1mm}
=~-igf^{mnl}\!\int\! dz \Big\{\big(\langle C^a_\mu(x)\barD^\xi C^{m\eta}(z)\rangle
-\langle C^a_\mu(x)\barD^\xi \tilC^{m\eta}(z)\rangle\big)
\langle C^n_\xi(z)\rangle_J^{[0]}\langle C^l_\eta(z)\rangle_J^{[0]}
\nonumber\\
&&\hspace{22mm}
+~\big(\langle C^a_\mu(x)C^n_\xi(z)\rangle
-\langle C^a_\mu(x)\tilC^n_\xi(z)\rangle\big)
(\langle \barD^\xi C^{m\eta}(z)\rangle_J^{[0]}-\xi\leftrightarrow\eta)
\langle C^l_\eta(z)\rangle_J^{[0]}\Big\}
\nonumber\\
&&\hspace{-1mm}
=~-igf^{mnl}\!\int\! dz \Big\{(x|{1\over \Box^{\mu\eta}+i\epsilon p_0}\bar{P}^\xi|z)^{am}
\langle C^n_\xi(z)\rangle_J^{[0]}\langle C^{l\eta}(z)\rangle_J^{[0]}
\nonumber\\
&&\hspace{22mm}
-~i(x|{1\over \Box^{\mu\xi}+i\epsilon p_0}|z)^{an}
(\langle \barD^\xi C^{m\eta}(z)\rangle_J^{[0]}-\xi\leftrightarrow\eta)
\langle C^l_\eta(z)\rangle_J^{[0]}\Big\}
\label{a17}
\end{eqnarray}
which is the desired result.

Similarly, in the $g^2$ order one obtains after some algebra
\begin{eqnarray}
&&\hspace{-1mm}
\langle C^a_\mu(x)\rangle_J^{[2]}
\label{a18}\\
&&\hspace{-1mm}
=~-igf^{mnl}\!\int\! dz \Big\{(x|{1\over \Box^{\mu\eta}+i\epsilon p_0}\bar{P}^\xi|z)^{am}
\Big[\langle C^n_\xi(z)\rangle_J^{[1]}\langle C^{l\eta}(z)\rangle_J^{[0]}
+\langle C^n_\xi(z)\rangle_J^{[0]}\langle C^{l\eta}(z)\rangle_J^{[1]}\Big]
\nonumber\\
&&\hspace{-1mm}
+~i(x|{1\over \Box^{\mu\xi}+i\epsilon p_0}|z)^{am}
\Big[(\langle \barD^\xi C^{n\eta}(z)\rangle_J^{[1]}-\xi\leftrightarrow\eta)
\langle C^l_\eta(z)\rangle_J^{[0]}
+(\langle \barD^\xi C^{m\eta}(z)\rangle_J^{[0]}-\xi\leftrightarrow\eta)
\nonumber\\
&&\hspace{-1mm}
\times~\langle C^l_\eta(z)\rangle_J^{[1]}\Big]\Big\}
+~g^2\int d^4z(x|{1\over \Box^{\mu\xi}+i\epsilon p_0}|z)^{am}f^{mnb}f^{cdn}
\langle C^{b\eta}(z)\rangle^{[0]}_J\langle C^{c\xi}(z)\rangle^{[0]}_J
\langle C^d_\eta(z)\rangle^{[0]}_J
\nonumber
\end{eqnarray}
At arbitrary order in $g$ the structure similar to Eq. (\ref{a18})
 can be proved by induction.
 
  Thus, we see that Eq. (\ref{a1}) is given by a set of Feynman
diagrams with retarded Green functions. In a similar way, one can demonstrate
that 
\begin{eqnarray}
&&\hspace{-1mm}
\int\! D\tilC DC~\tilC^a_\mu(x)~\exp\Big\{\!\int\! dz~\Big[{i\over 2}\tilC^{m\xi}\Box^{mn}_{\xi\eta}\tilC^{n\eta}
\nonumber\\
&&\hspace{-1mm}
+~igf^{mnl}\barD^\xi \tilC^{m\eta} \tilC^n_\xi \tilC^l_\eta 
+\frac{ig^2}{4}f^{abm}f^{cdm}\tilC^{a\xi} \tilC^{b\eta}\tilC^c_\xi \tilC^d_\eta
-iJ^m_\xi\tilC^{m\xi}
\nonumber\\
&&\hspace{-1mm}
-~
{i\over 2}C^{m\xi}\Box^{mn}_{\xi\eta}C^{n\eta}
-igf^{mnl}\barD^\xi C^{m\eta} C^n_\xi C^l_\eta 
-{ig^2\over 4}f^{abm}f^{cdm}C^{a\xi} C^{b\eta}C^c_\xi C^d_\eta+iJ^m_\xi C^{m\xi}\Big]\Big\}
\nonumber\\
&&\hspace{-1mm}
=~{\rm r.h.s.~ of~Eq. ~(\ref{a1})}~=~\langle C^a_\mu(x)\rangle_J~
\label{a19}
\end{eqnarray}

\section{Appendix B}
To find matrix $\Omega(x)$ satisfying Eqs. (\ref{omegasy}) we will solve the following auxiliary problem: 
we fix $x_\perp$ as a parameter and find the solution of Yang-Mills equations 
\begin{equation}
\cald^\nu \calf^a_{\mu\nu}(x_\ast,x_\bu)=~0
\label{kleq2d}
\end{equation}
in 2-dimensional gluodynamics with initial conditions
\begin{eqnarray}
&&\hspace{-11mm}
\cala_\mu(x_\ast,x_\bu)\stackrel{x_\ast\rightarrow -\infty}{=}\barA_\mu(x_\bu),~~~~
\cala_\mu(x_\ast,x_\bu)\stackrel{x_\bu\rightarrow -\infty}{=}\barB_\mu(x_\ast)
\label{inicondi2d}
\end{eqnarray}
Since 2-dimensional gluodynamics is a trivial theory, the solution of the equation (\ref{kleq2d})
will be a pure-gauge field $\cala_\mu=\Omega i\partial_\mu\Omega^\dagger$ with $\Omega(x_\ast,x_\bu)$ being 
the sought-for matrix satisfying Eqs. (\ref{omegasy}).

Let us first demonstrate that the solution $\cala_\mu(x_\ast,x_\bu)$ of the YM equations (\ref{kleq2d}) with boundary 
conditions (\ref{inicondi2d}) in two longitudinal dimensions is a pure gauge. 
To this end, we will construct 
$\cala_\mu(x_\ast,x_\bu)$ order by order in perturbation theory (see Fig. \ref{fig:3}, but now in two dimensions) and prove that $F^a_{\mu\nu}(\cala)~=~0$.

We are looking for the solution of Eq. (\ref{kleq2d}) in the form
\begin{equation}
\cala_\mu(x_\ast,x_\bu)~=~\barA_\mu(x_\ast,x_\bu)+\barC_\mu(x_\ast,x_\bu),~~~~~ \barA_\ast(x_\ast,x_\bu)~=~\barA_\ast(x_\bu),~~\barA_\bu(x_\ast,x_\bu)~=~\barB_\bu(x_\ast)
\end{equation}
Imposing the background-gauge condition
\begin{equation}
\barD^\mu \bar{C}_\mu(x_\ast,x_\bu)~=~0
\label{bfcond}
\end{equation}
we get the equation
\begin{equation}
\hspace{-1mm}
(\barP^2 g_{\mu\nu}+2ig\barF_{\mu\nu})^{ab}\barC^{b\nu}~=~\barD^{ab\xi}\barF^b_{\xi\mu}+
gf^{abc}(2\barC^b_\nu \barD^\nu \barC^c_\mu
-\barC^b_\nu \barD_\mu \barC^{c\nu})
-g^2f^{abm}f^{cdm}\barC^{b\nu}\barC^c_\mu\barC^d_\nu
\label{klypec}
\end{equation}
where $\barD_\mu\equiv (\partial_\mu-ig[\barA_\mu,)$ and $\barF_{\ast\bu}~=~-ig[\barA_\ast,\barB_\bu]$.  The boundary 
conditions (\ref{inicondi2d}) in terms of $C$ fields read
\begin{equation}
\hspace{-1mm}
C_\mu(x_\ast,x_\bu)\stackrel{x_\ast\rightarrow -\infty}{=}0,~~~~
C_\mu(x_\ast,x_\bu)\stackrel{x_\bu\rightarrow -\infty}{=}0
\label{inicondi2dc}
\end{equation}
It is convenient to rewrite the equation (\ref{klypec})  in components as
\begin{eqnarray}
&&\hspace{-1mm}
2(\barP_\bu\barP_\ast)^{ab} \barC_\bu^b~
\label{klypecomp}\\
&&\hspace{-1mm}
=~\barD^{ab}_\bu\barF^b_{\ast\bu}+ig\barF_{\ast\bu}^{ab}\barC_\bu^b
+g\barD^{aa'}_\bu(f^{a'bc}\barC^b_\ast\barC^c_\bu)
+2gf^{abc}\barC^b_\bu \barD_\ast \barC^c_\bu
-g^2f^{abm}f^{cdm}\barC^b_\bu\barC^c_\bu\barC^d_\ast
\nonumber\\
&&\hspace{-1mm}
2(\barP_\ast\barP_\bu)^{ab} \barC_\ast^b~
\nonumber\\
&&\hspace{-1mm}
=~-\barD^{ab}_\ast\barF^b_{\ast\bu}-ig\barF_{\ast\bu}^{ab}\barC_\ast^b
-g\barD^{aa'}_\ast(f^{a'bc}\barC^b_\ast\barC^c_\bu)
+2gf^{abc}\barC^b_\ast \barD_\bu \barC^c_\ast
-g^2f^{abm}f^{cdm}\barC^b_\ast\barC^c_\ast\barC^d_\bu
\nonumber
\end{eqnarray}
We will solve this equation by iterations in $\barF_{\ast\bu}$ and prove that $\calf_{\ast\bu}~=~0$ in all orders. 

In the first order we get the equation
\begin{equation}
\hspace{-1mm}
2(\barP_\bu\barP_\ast)^{ab} \barC_\bu^b~=~\barD^{ab}_\bu\barF^b_{\ast\bu},~~~~~2(\barP_\ast\barP_\bu)^{ab} \barC_\ast^b~=~-\barD^{ab}_\ast\barF^b_{\ast\bu}
\label{klypec1}
\end{equation}
The solution satisfying boundary conditions (\ref{inicondi2dc}) has the form
\begin{eqnarray}
\hspace{-1mm}
\barC^{(1)}_\bu~=~-{i/2\over \barP_\ast+i\epsilon}\barF_{\ast\bu}~~&\Leftrightarrow&~~
\barC^{(1)a}_\bu(x)~=~-{i\over 2}\!\int\! d^2z_\parallel
(x|{1\over \barP_\ast+i\epsilon}|z)^{ab} \barF_{\ast\bu}^b(z)
\nonumber\\
\hspace{-1mm}
\barC^{(1)}_\ast~=~{i/2\over \barP_\bu+i\epsilon}\barF_{\ast\bu}~~\hspace{4mm}&\Leftrightarrow&~~
\barC^{(1)a}_\ast(x)~=~{i\over 2}\!\int\! d^2z_\parallel
(x|{1\over\barP_\bu +i\epsilon}|z)^{ab} \barF_{\ast\bu}^b(z) \hspace{4mm}
\label{ce1}
\end{eqnarray}
Using the explicit form of the propagators in external $\barA_\ast$ and $\barB_\bu$ fields
\begin{eqnarray}
&&\hspace{-1mm}
(x|{1\over \barP_\bu +i\epsilon}|z)~=~-i\delta(x_\bu-z_\bu)\theta(x_\ast-z_\ast)[x_\ast,z_\ast]^{\barB_\bu}
\nonumber\\
&&\hspace{-1mm}
(x|{1\over \barP_\ast +i\epsilon}|z)~=~-i\delta(x_\ast-z_\ast)\theta(x_\bu-z_\bu)[x_\bu,z_\bu]^{\barA_\ast}
\label{props2d}
\end{eqnarray}
we get $\barC^{(1)}$ in the form
\begin{eqnarray}
&&\hspace{-11mm}
\barC^{(1)}_\ast(x)~=~-{i\over s}\!\int_{-\infty}^{x_\ast}\! dz_\ast 
~[x_\ast,z_\ast]^{A_\bu}[\barA_\ast(x_\bu),\barA_\bu(z_\ast)][z_\ast,x_\ast]^{A_\bu}
\nonumber\\
&&\hspace{-11mm}
\barC^{(1)}_\bu(x)~=~{i\over s}\!\int_{-\infty}^{x_\bu}\! dz_\bu 
~[x_\bu,z_\bu]^{A_\ast}[\barA_\ast(z_\bu),\barA_\bu(x_\ast)][z_\bu,x_\bu]^{A_\ast}
\label{ce1expl}
\end{eqnarray}
From this equation it is clear that $C_\mu^{(1)}(x_\ast,x_\bu)$ vanishes if $x_\ast\rightarrow -\infty$ and/or $x_\bu\rightarrow -\infty$ 
(recall that we assume $\barA_\ast(x_\bu)\stackrel{x_\bu\rightarrow\pm\infty}{\rightarrow}0$
and $\barB_\bu(x_\ast)\stackrel{x_\ast\rightarrow\pm\infty}{\rightarrow}0$).

Also, form Eq. (\ref{ce1}) we see that
\begin{equation}
\hspace{-1mm}
\barD_\ast\barC^{(1)}_\bu~=~-\half\barF_{\ast\bu},~~~~~\barD_\bu\barC^{(1)}_\ast~=~\half\barF_{\ast\bu}
\label{demucemu1}
\end{equation}
and therefore
\begin{equation}
\hspace{-1mm}
\calf_{\ast\bu}~=~\barF_{\ast\bu}+\barD_\ast\barC^{(1)}_\bu-\barD_\bu\barC^{(1)}_\ast+O(\barF^2)~=~O(\barF^2)
\label{vanish1}
\end{equation}
so in the first order in $\barF$ the field strength of the solution of classical equation (\ref{klypec}) vanishes. 

In the second order the equations for the field $C_\mu$ take the form
\begin{eqnarray}
&&\hspace{-1mm}
2(\barP_\bu\barP_\ast)^{ab} \barC_\bu^{(2)b}~=g\barD^{aa'}_\bu(f^{a'bc}\barC^{(1)b}_\ast\barC^{(1)c}_\bu)
~~\Rightarrow~~\barC_\bu^{(2)a}
~=-{ig\over 2}\big({1\over \barP_\ast+i\epsilon}\big)^{aa'}f^{a'bc}\barC^{(1)b}_\ast\barC^{(1)c}_\bu
\nonumber\\
&&\hspace{-1mm}
2(\barP_\ast\barP_\bu)^{ab} \barC_\ast^{(2)b}~=
-g\barD^{aa'}_\ast(f^{a'bc}\barC^{(1)b}_\ast\barC^{(1)c}_\bu)
~~\Rightarrow~~\barC_\ast^{(2)a}~={ig\over 2}\big({1\over \barP_\bu+i\epsilon}\big)^{aa'}
f^{a'bc}\barC^{(1)b}_\ast\barC^{(1)c}_\bu
\nonumber\\
\label{cedva}
\end{eqnarray}
where we used Eq. (\ref{demucemu1}) to reduce the r.h.s. Again, from the explicit form of the propagators (\ref{props2d})
we get
\begin{eqnarray}
&&\hspace{-1mm}
\barC^{(2)}_\ast(x)~=~-{ig\over s}\!\int_{-\infty}^{x_\ast}\! dz_\ast 
~[x_\ast,z_\ast]^{A_\bu}[\barC^{(1)}_\ast(z_\ast,x_\bu),\barC^{(1)}_\bu(z_\ast,x_\bu)][z_\ast,x_\ast]^{A_\bu}
\nonumber\\
&&\hspace{-1mm}
\barC^{(2)}_\bu(x)~=~{ig\over s}\!\int_{-\infty}^{x_\bu}\! dz_\bu 
~[x_\bu,z_\bu]^{A_\ast}[\barC^{(1)}_\ast(x_\ast,z_\bu),\barC^{(1)}_\bu(x_\ast,z_\bu)][z_\bu,x_\bu]^{A_\ast}
\label{ce2expl}
\end{eqnarray}
from which it is clear that $\barC^{(2)}_\mu$ satisfy boundary conditions (\ref{inicondi2dc}) 
(recall that we already proved that $\barC^{(1)}_\mu$ satisfy  Eq. (\ref{inicondi2dc})). Next, we use 
\begin{eqnarray}
&&\hspace{-11mm}
\barD_\ast\barC^{(2)a}_\bu~=~-{g\over 2} f^{abc}\barC^{(1)b}_\ast\barC^{(1)c}_\bu,~~~~~~~~~~~~
\barD_\bu\barC^{(2)a}_\ast~=~{g\over 2}f^{abc}\barC^{(1)b}_\ast\barC^{(1)c}_\bu
\label{demucemu2}
\end{eqnarray}
to prove that $\calf_{\ast\bu}$ vanishes in the second order:
\begin{eqnarray}
&&\hspace{-1mm}
\calf_{\ast\bu}^{a}~=~F^a_{\ast\bu}(\barA+C^{(1)}+C^{(2)})+O(\barG^3)
\nonumber\\
&&\hspace{-1mm}=~\barF^a_{\ast\bu}
+(\barD_\ast \barC^{(1)}_\bu-\barD_\bu \barC^{(1)}_\ast)^a
+(\barD_\ast \barC^{(2)}_\bu-\barD_\bu \barC^{(2)}_\ast)^a+gf^{abc}\barC^{(1)b}_\ast\barC^{(1)c}_\bu+O(\barG^3)~
\nonumber\\
&&\hspace{-1mm}=~O(\barG^3)
\label{vanish2}
\end{eqnarray}

In the third order we get
\begin{eqnarray}
&&\hspace{-1mm}
2(\barP_\bu\barP_\ast)^{ab} \barC_\bu^{(3)b}~=~g\barD^{aa'}_\bu f^{a'bc}(\barC^{(1)b}_\ast\barC^{(2)c}_\bu+\barC^{(2)b}_\ast\barC^{(1)c}_\bu)
\nonumber\\
&&\hspace{-1mm}
2(\barP_\ast\barP_\bu)^{ab} \barC_\ast^{(3)b}~=~
-g\barD^{aa'}_\ast f^{a'bc}(\barC^{(1)b}_\ast\barC^{(2)c}_\bu+\barC^{(2)b}_\ast\barC^{(1)c}_\bu)
\label{cetri}
\end{eqnarray}
where again we used Eqs. (\ref{demucemu1}) and (\ref{demucemu2}) to reduce the r.h.s.
The solution is 
\begin{eqnarray}
&&\hspace{-1mm}
\barC_\bu^{(3)a}
~=~-{ig\over 2}\big({1\over \barP_\ast+i\epsilon}\big)^{aa'}f^{a'bc}(\barC^{(1)b}_\ast\barC^{(2)c}_\bu+\barC^{(2)b}_\ast\barC^{(1)c}_\bu)
\nonumber\\
&&\hspace{-1mm}
\barC_\ast^{(3)a}~=~{ig\over 2}\big({1\over \barP_\bu+i\epsilon}\big)^{aa'}
f^{a'bc}(\barC^{(1)b}_\ast\barC^{(2)c}_\bu+\barC^{(2)b}_\ast\barC^{(1)c}_\bu)
\label{salushen}
\end{eqnarray}
Again, from the explicit form of propagators (\ref{props2d}) it is clear that $\barC_\mu^{(3)}$ satisfy boundary conditions (\ref{inicondi2d}) if $\barC_\mu^{(1)}$ and $\barC_\mu^{(2)}$ 
do (which we already proved). Next, from
\begin{equation}
\barD_\ast\barC_\bu^{(3)a}~=~-{g\over 2}f^{abc}(\barC^{(1)b}_\ast\barC^{(2)c}_\bu+\barC^{(2)b}_\ast\barC^{(1)c}_\bu),~~~
\barD_\bu\barC_\ast^{(3)a}~=~{g\over 2}f^{abc}(\barC^{(1)b}_\ast\barC^{(2)c}_\bu+\barC^{(2)b}_\ast\barC^{(1)c}_\bu)
\label{demucemu3}
\end{equation}
we see that $\calf_{\ast\bu}$ vanishes in the third order:
\begin{eqnarray}
&&\hspace{-1mm}
\calf^a_{\ast\bu}~=~F_{\ast\bu}^a(\barA+\barC^{(1)}+\barC^{(2)}+\barC^{(3)})+O(\barG^4)
\nonumber\\
&&\hspace{-1mm}
=~\barG^a_{\ast\bu}+(\barD_\ast \barC^{(1)}_\bu-\barD_\bu \barC^{(1)}_\ast)^a
+(\barD_\ast \barC^{(2)}_\bu-\barD_\bu \barC^{(2)}_\ast)^a+(\barD_\ast \barC^{(3)}_\bu-\barD_\bu \barC^{(3)}_\ast)^a
\nonumber\\
&&\hspace{-1mm}
+~gf^{abc}(\barC^{(1)b}_\ast\barC^{(1)c}_\bu+\barC^{(1)b}_\ast\barC^{(2)c}_\bu+\barC^{(2)b}_\ast\barC^{(1)c}_\bu)+O(\barG^4)~=~O(\barG^4)
\label{vanish3}
\end{eqnarray}
 Note also that Eqs. (\ref{demucemu1}),  (\ref{demucemu2}) and 
(\ref{demucemu3}) illustrate self-consistency check for the background-field condition (\ref{bfcond}).

One can continue and prove by induction that $\calf_{\ast\bu}$ vanishes in an arbitrary order in $\barG_{\ast\bu}^n$ and therefore 
the field $\cala_\mu$ is a pure gauge
\begin{eqnarray}
&&\hspace{-1mm}
\cala_\ast(x_\ast,x_\bu)~=~\barA_\ast(x_\bu)+\barC_\ast(x_\ast,x_\bu)~=~\Omega(x_\ast,x_\bu)i\partial_\ast\Omega^\dagger(x_\ast,x_\bu)~~~~
\nonumber\\
&&\hspace{-1mm}
\cala_\bu(x_\ast,x_\bu)~=~\barB_\bu(x_\ast)+\barC_\bu(x_\ast,x_\bu)~=~\Omega(x_\ast,x_\bu)i\partial_\bu\Omega^\dagger(x_\ast,x_\bu)
\label{pure1}
\end{eqnarray}

Now we shall demonstrate that the matrix $\Omega$ satisfies our requirement (\ref{omegasy}). Since $C_\ast(x_\ast\rightarrow -\infty,x_\bu)~=~0$ due to Eq. (\ref{inicondi2d}), we get
\begin{equation}
\Omega(-\infty,x_\bu)i\partial_\ast\Omega^\dagger(-\infty,x_\bu)~=~\barA_\ast(x_\bu)~~~\Rightarrow~~~\Omega(-\infty,x_\bu)~=~[x_\bu,-\infty_\bu]^{\barA_\ast}
\label{pure2}
\end{equation}
Similarly,
\begin{equation}
\Omega(x_\ast,-\infty)i\partial_\bu\Omega^\dagger(x_\ast,-\infty)~=~\barB_\bu(x_\ast)~~~\Rightarrow~~~\Omega(x_\ast,-\infty)~=~[x_\ast,-\infty_\ast]^{\barB_\bu}
\label{pure3}
\end{equation}
One can also construct the expansion of matrix $\Omega$ in powers of $\barA_\ast$ and $\barB_\bu$. For example, up to the fifth power of the $\barA_\mu$ fields
\begin{eqnarray}
&&\hspace{-1mm}
\Omega(x_\ast,x_\bu)~
\label{omega}\\
&&=~\half\{[x_\ast,-\infty_\ast]^{\barB_\bu},[x_\bu,-\infty_\bu]^{\barA_\ast}\}
-{1\over 4}\big(\big[[x_\bu,-\infty_\bu]^{\barA_\ast},[x_\ast,-\infty_\ast]^{\barB_\bu}\big]\big)^2
\nonumber\\
&&-~{4g^4\over s^4}\!\int_{-\infty}^{x_\ast}\!dx'_\ast\!\int_{-\infty}^{x'_\ast}\! dx''_\ast
\!\int_{-\infty}^{x_\bu}\!dx'_\bu\!\int_{-\infty}^{x'_\bu}\! dx''_\bu~
\big[[\barA_\bu(x'_\ast),\barA_\ast(x'_\bu)],[\barA_\bu(x''_\ast),\barA_\ast(x''_\bu)]\big]
\nonumber
\end{eqnarray}
Now, for each $x_\perp$ we solve auxiliary 2-dimensional classical problem (\ref{kleq2d}) and find\\ $\Omega(x_\ast,x_\bu,x_\perp)$ satisfying the requirement (\ref{omegasy}).

\bibliography{fact1}
\bibliographystyle{JHEP}

\end{document}